\documentclass[lettersize,journal]{IEEEtran}
\usepackage{amsmath,amsfonts}
\usepackage{algorithmic}
\usepackage{algorithm}
\usepackage{array}
\usepackage[caption=false,font=normalsize,labelfont=sf,textfont=sf]{subfig}
\usepackage{textcomp}
\usepackage{stfloats}
\usepackage{url}
\usepackage{verbatim}
\usepackage{graphicx}
\usepackage{cite}
\usepackage{listings}
\usepackage{amsmath,amsfonts}
\usepackage{graphicx}
\usepackage{amsfonts}
\usepackage{array}
\usepackage{amssymb}
\usepackage{helvet}
\usepackage{color}
\usepackage{bm}
\usepackage{float}
\usepackage{cite}
\usepackage{multirow}
\usepackage{slashbox}
\usepackage{amssymb}
\usepackage[caption=false,font=normalsize,labelfont=sf,textfont=sf]{subfig}
\usepackage{amsmath,amsfonts}
\usepackage{algorithmic}
\usepackage{algorithm}
\usepackage{textcomp}
\usepackage{stfloats}
\usepackage{url}
\usepackage{verbatim}
\usepackage{graphicx}
\usepackage{cite}
\usepackage{amssymb}
\usepackage{makecell}

\newcommand{\RNum}[1]{\uppercase\expandafter{\romannumeral #1\relax}}

\begin{document}

\title{YOLO: An Efficient Terahertz Band Integrated Sensing and Communications Scheme with \\ Beam Squint}

\author{Hongliang Luo, Feifei Gao,~\IEEEmembership{Fellow,~IEEE,} Hai Lin,~\IEEEmembership{Senior Member,~IEEE,} \\ Shaodan Ma,~\IEEEmembership{Senior Member,~IEEE,} and H. Vincent Poor,~\IEEEmembership{Life Fellow,~IEEE}
\thanks{This work has been accepted by IEEE Transactions on Wireless Communications (TWC).}
\thanks{Manuscript received 16 September 2023; revised 12 November 2023 and 19 January 2024; accepted 27 January 2024; date of current version 2  February 2024. 
The  editor  coordinating the review of this article and approving it for publication is Prof. Kaiming Shen. 
This work was supported in part  by the National Natural Science Foundation of China under Grants  {62325107, 62341107, U23A20272, and  62261160650}, 
in part by the Japan Society for the 
Promotion of Science (JSPS) Grants-in-Aid for Scientific Research 
(KAKENHI) under Grant 22H01491, 
in part by the Science and Technology Development Fund, Macau SAR under Grants 0087/2022/AFJ and SKL-IOTSC(UM)-2024-2026, 
in part by the Research Committee of University of Macau under Grants MYRG-GRG2023-00116-FST-UMDF and MYRG2020-00095-FST,
and in part by the U.S National Science Foundation under Grants CNS-2128448 and ECCS-2335876. 
(\textit{Corresponding author: Feifei Gao.})}
\thanks{H. Luo and F. Gao are with 
Department of Automation, Tsinghua University,
Beijing, P.R. China (email: luohl23@mails.tsinghua.edu.cn, feifeigao@ieee.org).}
\thanks{Hai Lin is with the Department of Electrical and Electronic Systems Engineering, Graduate School of Engineering, Osaka Metropolitan University, Sakai, Osaka, Japan (e-mail: hai.lin@ieee.org).}
\thanks{Shaodan Ma is with the State Key Laboratory of Internet of Things for Smart City and the Department of Electrical and Computer Engineering, University of Macau, Macao S.A.R., P.R. China (e-mail: shaodanma@um.edu.mo).}
\thanks{H. Vincent Poor is with the Department of Electrical and Computer Engineering, Princeton University, Princeton, NJ USA (e-mail: poor@princeton.edu).}
}

\maketitle

\begin{abstract}
Using communications signals for dynamic target sensing is an important  component of integrated sensing and communications (ISAC).
In this paper, we propose to utilize the
 beam squint effect  to realize fast non-cooperative dynamic target sensing in massive
multiple input and multiple output
 (MIMO) Terahertz band communications systems.
Specifically,
we construct a wideband channel model of the  echo signals,
and design a  beamforming strategy that controls the range of beam squint by adjusting the values of phase shifters and true time delay lines.
With this design, beams at  different subcarriers can be aligned  along different directions in a planned way.
Then 
the received  echo signals at different subcarriers will carry target information in different directions, based on which  the targets' angles can be estimated through  sophisticatedly designed algorithm.
Moreover,
we propose a supporting method based on extended array signal estimation, which utilizes the phase changes of different frequency subcarriers within different orthogonal frequency division multiplexing (OFDM) symbols to estimate the distances and velocities of dynamic targets.
Interestingly,
the proposed sensing scheme only needs  to transmit and receive the signals once, which can be termed as \emph{You Only Listen Once} (YOLO).
Compared with the traditional ISAC methods that require time consuming beam sweeping, the proposed one  greatly reduces the sensing overhead.
Simulation results are provided to  demonstrate the effectiveness of the proposed schemes.
\end{abstract}


\section{Introduction}

Integrated sensing and communications (ISAC) is anticipated to be a key  technology of  sixth generation (6G) mobile communications \cite{ISAC20231,9397776,ISAC1}.
Its basic idea is to use communication signals to sense various aspects of the real  physical  world, such as  target's position, architectural composition,   human activity, etc.
Once the ISAC system obtains such information, it can not only enhance the data transmission performance\cite{d2}, but also support a variety of intelligent applications, such as connected vehicles and smart factory\cite{butalso1,9040264}.

 As the decisive function of ISAC, the ultimate functionality of sensing
 is to construct a mapping relationship from the  \emph{real physical world} to the  \emph{digital twin world}.
Note that the real physical world is composed of\textbf{ static environments} (such as roads and buildings) and \textbf{dynamic targets} (such as pedestrians and vehicles).
The  changes of  static environment are usually slow, and thus
 one can apply  various environmental reconstruction technologies to sense the static environment\cite{9727176,9516898,10182348}.
 However,
 since the changes of dynamic targets are rapid, it is necessary to update the parameters of dynamic targets in real-time.
Generally, the\textbf{ dynamic targets sensing problem} can be classified into two categories:
(1) sensing  cooperative communications users, e.g., mobile phones;  and
(2) sensing  non-cooperative dynamic targets that are not communicating with the  base station (BS), e.g., moving objects or users not in communications status.

 In  massive multiple input and multiple output (MIMO) arrays based ISAC systems,
 a series of signal processing algorithms can be implemented for  sensing  cooperative users.
For example, by processing the pilot signals of the communications system, the time of arrival (TOA),   and the angle of departure (AOD), etc.,  can be extracted to estimate a user's position\cite{g333,g111,g222,9506874}.
The sensing of non-cooperative  targets has traditionally been accomplished via radar systems\cite{202301041,8425968,2bs3}.
Currently, there are many works focusing  on non-cooperative target sensing under   ISAC framework. 
For example, 
Z.~Gao~\emph{et.~al.}  proposed an ISAC processing framework relying on  MIMO systems,
which applied  compressed sampling   to facilitate
target sensing and other  processing\cite{9898900}.
Z. Wang \emph{et.~al.} proposed a simultaneously transmitting and reflecting surface  enabled ISAC framework, in which the two-dimensional maximum likelihood estimation was utilized to estimate the direction-of-arrival of the sensing target\cite{10050406}.
C.~Mush \emph{et.~al.}  studied the potential applications of autoencoders in ISAC systems and 
presented  a novel perspective on target detection\cite{2023arXiv230109439M}.
 However, these works  consider only the sensing of static targets,
which entails difficulty in practical scenarios where static targets are coupled
  with  static environments and are difficult to separate.
In terms of non-cooperative dynamic target sensing,
 Y. Li \emph{et. al.} proposed a two-stage algorithm using  orthogonal frequency division multiplexing (OFDM) signals to estimate the parameters of  targets\cite{8918315}. 
P. Kumari \emph{et.~al.} proposed  an ISAC system   for the internet of vehicles, which realized vehicle to vehicle  communications and full duplex radar sensing\cite{8114253}. X.~Chen \emph{et. al.}  proposed a multiple signal classification  based ISAC system that can attain high estimation accuracy for dynamic target sensing\cite{10048770}.
Z.~Wei \emph{et.~al.} proposed the use of the iterative two-dimensional  fast Fourier transform  and iterative cyclic cross-correlation  methods to realize short distance and long distance dynamic target sensing, respectively\cite{10011549}.
Z.~Han~\emph{et.~al.} designed  a novel multistatic MIMO-ISAC system in cellular networks, and   made the use of widespread BSs to perform  dynamic target sensing over a wide area\cite{2023arXiv230512994H}.
Nevertheless, although these approaches consider the sensing of dynamic targets, they do not address the serious
interference caused by static environmental clutter on dynamic target sensing, thus limiting the sensing performance in some practical scenarios. 

\begin{figure*}[!t]
\centering
\includegraphics[width=140mm]{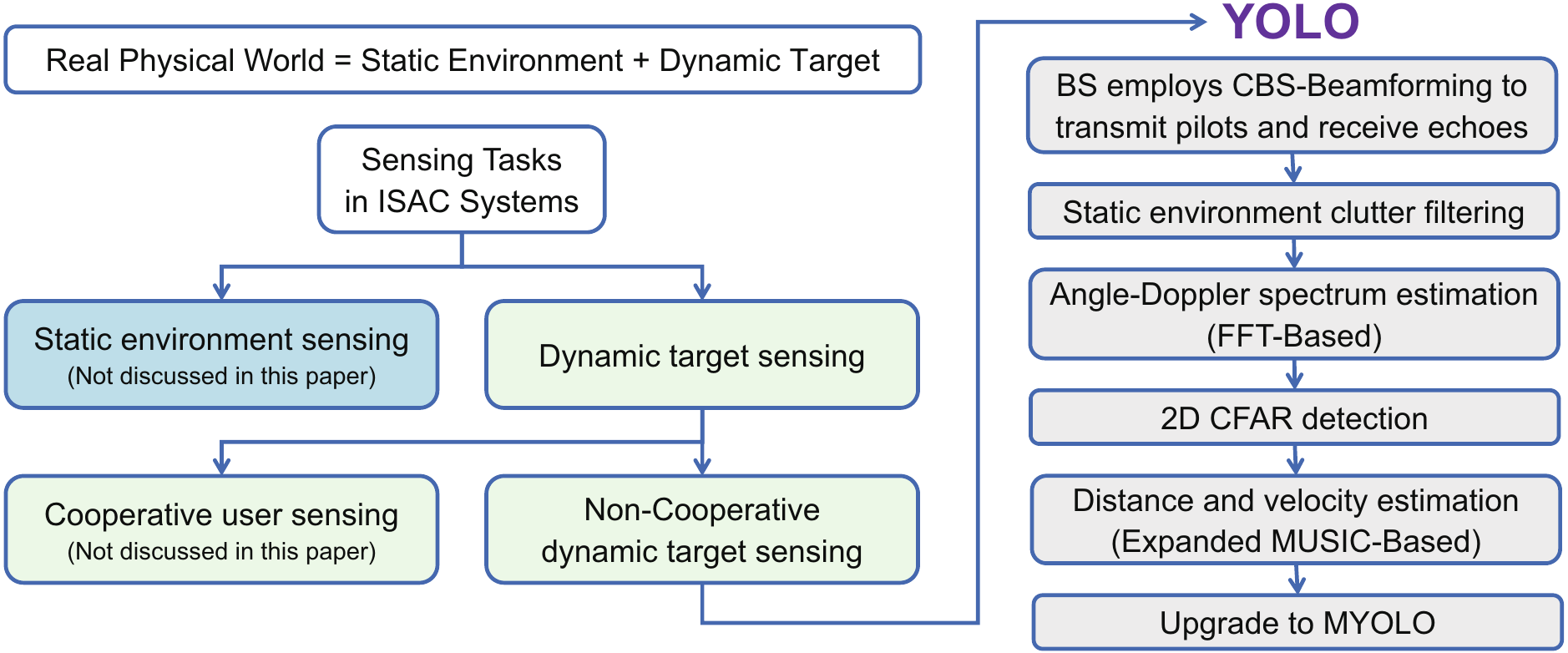}
\caption{ISAC sensing task classification and  flowchart of the proposed YOLO sensing scheme.}
\label{fig_1}
\end{figure*}

In addition, the aforementioned algorithms for dynamic target sensing consider only  narrowband signals,
while  wideband signals from mmWave and Terahertz frequency bands  are expected to  dominate in 6G communications\cite{8732419,9882323}.
Specially, Terahertz communications  is recognized  as a highly promising technology for the 6G and  beyond era, due to its unique potential to support  terabit-per-second transmission in emerging applications\cite{9887921,9766110}.
Meanwhile, sensing in the Terahertz band with MIMO can achieve higher accuracy thanks to the corresponding  high directivity and high temporal resolution\cite{9782674,9013639}.
Nevertheless, it is  by no means straightforward or may even be infeasible  to extend  narrowband sensing algorithms to  wideband scenarios.
For example,
 \cite{wblsp} and  \cite{my1} showed that for a wideband Terahertz  system,
the   \emph{beam squint} phenomenon would appear, in which the beams from different subcarriers would disperse to different directions, making beam directions  of some subcarriers deviate from the desired  direction.
Beam squint  is usually considered to be a negative effect for communications   and  should be mitigated with the aid of  true time delay lines\cite{7841766,b1,nn1}.
Interestingly, some  works have recently tried to enhance the sensing performance  through reverse utilization of beam squint. For example, the authors of  \cite{10058989} proposed a  user angle  sensing scheme with low pilot overhead based on beam squint and beam split, but they did not consider  distance and velocity estimation or  more general non-cooperative dynamic target sensing tasks.

In this paper, we  propose an efficient  non-cooperative dynamic target sensing
scheme
for Terahertz ISAC systems 
 with the aid of the  beam squint effect, in which the BS can quickly sense the dynamic target's position and velocity through  frequency-domain estimation.
The contributions of this paper are summarized as follows.


\begin{itemize}

\item  We consider, for the first time, both the static environment and the dynamic targets in the ISAC scenario, and derive a corresponding wideband echo channel model  in the time-frequency domain.

\item We analyze the beam squint effect in wideband ISAC systems and propose a \emph{controllable beam squint} strategy to assist   sensing, which is termed \emph{CBS-Beamforming}.

\item We  utilize the controllable beam squint to realize   fast sensing of non-cooperative dynamic targets, in which the BS only needs to transmit and receive the OFDM signal once to obtain the positions and velocities of all targets in the entire  area.
This scheme is termed
\emph{you only listen once} (YOLO), and a flowchart of it is presented  in Fig.~1.

\item We improve the accuracy of sensing by performing YOLO multiple times with different beam squint ranges, which technique is termed  
\emph{Multiple YOLO} (MYOLO).

\item We provide various simulations to demonstrate  the effectiveness of the proposed schemes.

\end{itemize}

The remainder of this paper is organized as follows.
In Section \RNum{2}, we introduce  the wideband  channel model involving both the static environment and the dynamic targets.
In Section~\RNum{3}, the beam squint effect is analyzed and
 the beamforming strategy based on controllable beam squint is derived.
Then we propose the dynamic target sensing algorithms
  in Section \RNum{4}.
Simulation results and conclusions are given in Section \RNum{5} and Section \RNum{6},
respectively.

\emph{Notation}:
Lower-case and upper-case boldface letters $\mathbf{a}$ and $\mathbf{A}$ denote vectors and matrices;
$\mathbf{a}^T$ and $\mathbf{a}^H$ denote the transpose and the conjugate transpose of  $\mathbf{a}$;
${\rm diag}(\mathbf{a})$ denotes a diagonal matrix with the diagonal elements constructed from $\mathbf{a}$;
$[\mathbf{a}]_n$  denotes the $n$-th element of  $\mathbf{a}$;
$[\mathbf{A}]_{i,j}$ denotes the $(i,j)$-th element of  $\mathbf{A}$; $\mathbf{A}[i_1:i_2,:]$ is the submatrix composed of all  elements in rows $i_1$ to $i_2$ of matrix $\mathbf{A}$;
$\mathbf{A}[:,j_1:j_2]$ is the submatrix composed of all  elements in columns $j_1$ to $j_2$ of matrix $\mathbf{A}$;
Symbol $\textbf{1}$ represents the all-ones matrix or vector with compatible dimensions;
$\left|\cdot\right|$  denotes the absolute operator;
and 
${\rm eig}(\cdot)$ represents the  eigenvalue decomposition function.

\section{System Model}

\begin{figure*}[!t]
\centering
\subfloat[]{\includegraphics[width=70mm]{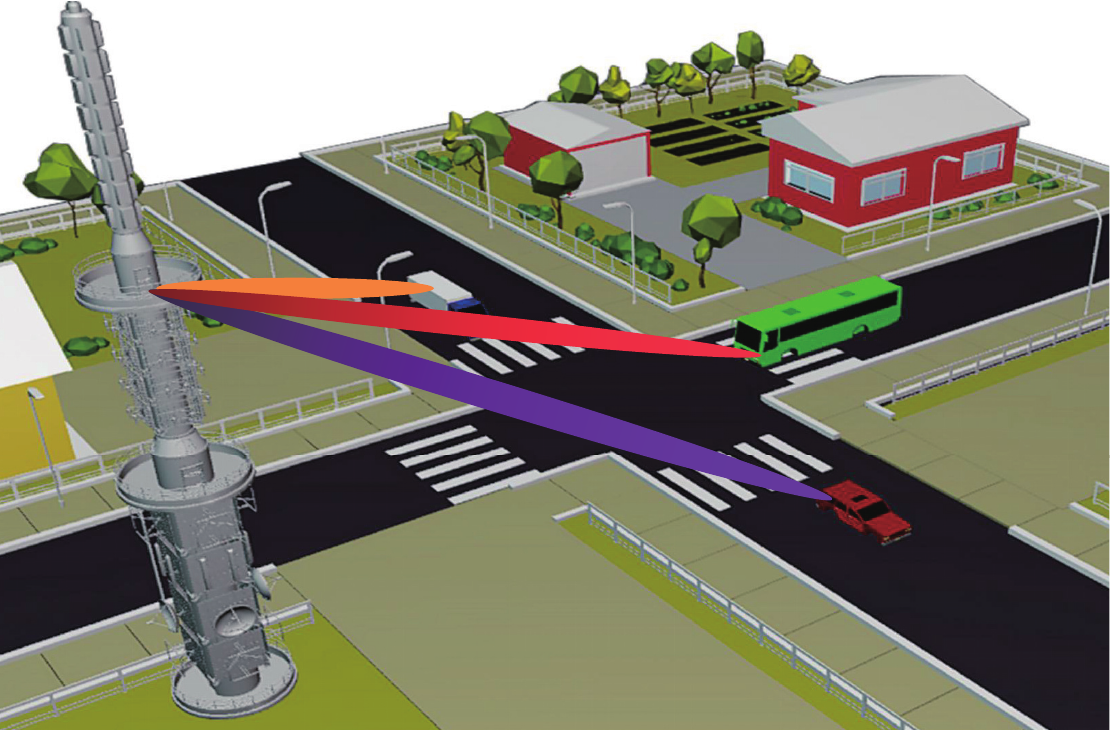}%
\label{fig_first_case}}
\hfil
\subfloat[]{\includegraphics[width=55mm]{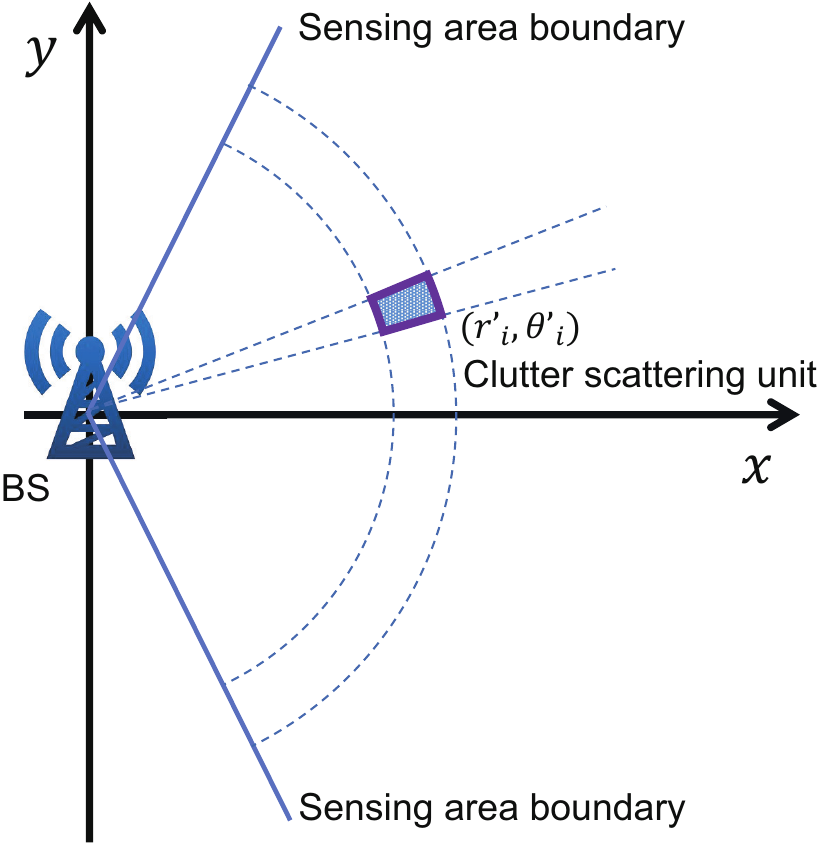}%
\label{fig_second_case}}
\caption{(a) System model.
(b) Schematic diagram of clutter scattering unit.}
\label{fig_sim}
\end{figure*}

Let us consider a massive MIMO system operating in the Terahertz frequency band with OFDM modulation.
The BS is equipped with a uniform linear array (ULA) of $N$ antennas, where the antenna spacing is $d\le \frac{\lambda}{2}$ and $\lambda$ is the wavelength.
All  antennas are located on the y-axis, and the position of the $n$-th antenna is $(0, nd)$,
where $n=-\frac{N-1}{2},...,\frac{N-1}{2}$.
Assume that the  BS employs a single radio frequency (RF) chain for sensing.
The carrier frequency and the transmission bandwidth are $f_c$ and $W$, respectively,
and thus the passband frequency range is $[f_c-\frac{W}{2},f_c+\frac{W}{2}]$.
Suppose there are $M+1$ subcarriers in total, where the $0$-th subcarrier has the lowest frequency $f_0=f_c-\frac{W}{2}$, and the $m$-th subcarrier frequency is $f_m=f_0+m\frac{W}{M}$. Let us denote the baseband frequency of the $m$-th subcarrier as $\widetilde{f}_m=m\frac{W}{M}$, where $m=0,1,2,...,M$.
Obviously, there is $f_m=f_0+\widetilde{f}_m$.
Assume that the BS works in full duplex mode and
has perfect self-interference cancellation\footnote{One can easily extend this work to  half duplex mode, in which  BS has   separate
transmitting  and  receiving  arrays with certain protection distance
 to eliminate  self-interference\cite{9898900}.
Nevertheless, we assume  full duplex operation here  mainly to illustrate the proposed YOLO scheme in a clearer  way.}.
We suppose that the BS enables an OFDM block containing $N_s$ consecutive OFDM symbols to realize dynamic target sensing, and the  symbol duration is 
$T_s = \frac{1}{\Delta f}=\frac{1}{W/M}$, where $\Delta f=\frac{W}{M}$ is the subcarrier frequency spacing.

As shown in Fig.~2(a), the real physical world scene is typically composed of static environment and dynamic targets.
When  the BS   sends the probing signal to the coverage area at the beginning of sensing,
 the receiving array will receive effective echoes caused by  dynamic targets of interest   (\textbf{dynamic target echo}) and undesired dense echoes caused by uninteresting  background environments.
 In radar systems, the undesired dense echoes  are usually referred to as ``\textbf{clutter}'', including ground clutter, sea clutter,  weather clutter, etc\cite{784056}.
 The key difference between dynamic target echoes and clutter lies in their different Doppler frequencies. That is, the Doppler frequency of a dynamic target echo is usually much higher than that of clutter, while the Doppler frequency of clutter is usually zero or a small non-zero value. Specifically, as the main component of urban environmental clutter, ground clutter is usually caused by static objects, such as land, mountains, roads,  buildings, etc, and its signal power intensity is usually about $50$ dB higher than that of the dynamic target echo, but its Doppler frequency is almost zero.
 Based on the above analysis, a practical modeling of echo signals should include both dynamic target echoes and static environment echoes, and the BS needs 
 to perform a series of signal processing  to  accurately detect and estimate the dynamic targets in a strong clutter background.
 However, some existing ISAC works\cite{9898900,10050406,2023arXiv230109439M} have not considered the objective existence of static environmental clutter and have instead  directly assumed that the stationary targets could be observed. Although some other ISAC works\cite{8918315,8114253,10048770,10011549,2023arXiv230512994H} have focused on dynamic target sensing, they have not considered the serious interference caused by static environmental clutter.

Here we model the clutter of  ISAC system based on a typical urban service scenario,
 in which the static environment can be divided into a finite number of grid units.
That is, we  divide the effective clutter area within the sensing range into equal intervals based on distance and angle dimensions, ultimately achieving a division of the static environment into a limited number of scattering units  as illustrated by the blue grid  in Fig.~2(b).
The size of each clutter scattering unit is determined by the angle resolution and distance resolution of the system.
Assume that the sensing range is divided into $I$ static clutter scattering units, where the position of the $i$-th static clutter scattering unit is denoted as $(r_i',\theta_i')$. Then the echo channel of this  unit on the $m$-th subcarrier of the $n_s$-th OFDM symbol can be modeled as
\begin{equation}
\begin{split}
\begin{aligned}
\label{deqn_ex1a}
\mathbf{H}_{i,n_s,m}' = \beta_i e^{-j2\pi f_m\frac{2r_i'}{c}}\mathbf{a}_{m}(\theta_i')\mathbf{a}_{m}^T(\theta_i') \in \mathbb{C}^{N\times N},
\end{aligned}
\end{split}
\end{equation}
where $\mathbf{a}_{m}(\theta) \in \mathbb{C}^{N\times 1}$ represents the steering vector of the antenna array
 towards the angle $\theta$ on the $m$-th subcarrier,  the $n$-th element in $\mathbf{a}_{m}(\theta)$ is $[\mathbf{a}_{m}(\theta)]_n=e^{j\frac{2\pi f_m }{c}nd\sin\theta}$, 
$\beta_i =\sqrt{\frac{\lambda^2}{(4\pi)^3 (r'_i)^4}}\sigma_{c,i}'$, 
and $\sigma_{c,i}'$ is the radar cross section (RCS) of the $i$-th  static clutter scattering unit.
The RCS of ground clutter can  be modeled using the Swerling $\rm \uppercase\expandafter{\romannumeral1}$ model\cite{9947033}, in which the probability density function (PDF) of the RCS $\sigma$ satisfies
\begin{equation}
\begin{split}
\begin{aligned}
\label{deqn_ex1a}
f(\sigma) = \frac{1}{\sigma_0} \exp(-\frac{\sigma}{\sigma_0}), \sigma \geq 0,
\end{aligned}
\end{split}
\end{equation}
where $\sigma_0$ is  the mean value of  the RCS.

In addition, we 
assume that there are $K$ dynamic targets within the sensing range of the BS.
The two-dimensional  position of the $k$-th dynamic target is $(x_k, y_k)$,  the corresponding polar coordinate  is $(r_k, \theta_k)$, and the radial velocity of the $k$-th dynamic target is $v_k$.
 Then the distance between the $k$-th  dynamic target and the $n$-th antenna of the BS is
\begin{equation}
\begin{split}
\begin{aligned}
\label{deqn_ex1a}
r_{k,n}&=\sqrt{x_k^2\!+\!(y_k\!-\!nd)^2}
\approx r_k - nd\sin\theta_k.
\end{aligned}
\end{split}
\end{equation}
Then the echo channel corresponding to the $k$-th dynamic  target on the $m$-th subcarrier of the $n_s$-th OFDM symbol is 
\begin{equation}
\begin{split}
\begin{aligned}
\label{deqn_ex1a}
\mathbf{H}_{k,n_s,m} = \alpha_k e^{j2\pi f_0 \frac{2v_k}{c}n_sT_s} e^{-j2\pi f_m\frac{2r_k}{c}}\mathbf{a}_{k,m}\mathbf{a}_{k,m}^T,
\end{aligned}
\end{split}
\end{equation}
where $\mathbf{a}_{k,m}\in \mathbb{C}^{N\times 1}$ represents the steering vector of the BS antennas with
the $n$-th element
$[\mathbf{a}_{k,m}]_n=e^{j\frac{2\pi f_m }{c}nd\sin\theta_k}$,
 $\alpha_k$ is  modeled as 
$\alpha_k =\sqrt{\frac{\lambda^2}{(4\pi)^3 r_k^4}}\sigma_{c,k}$, and 
 $\sigma_{c,k}$ 
 is the RCS of the $k$-th dynamic target, also following  the 
Swerling $\rm \uppercase\expandafter{\romannumeral1}$ model.

\begin{figure*}[!t]
\centering
\subfloat[]{\includegraphics[width=80mm]{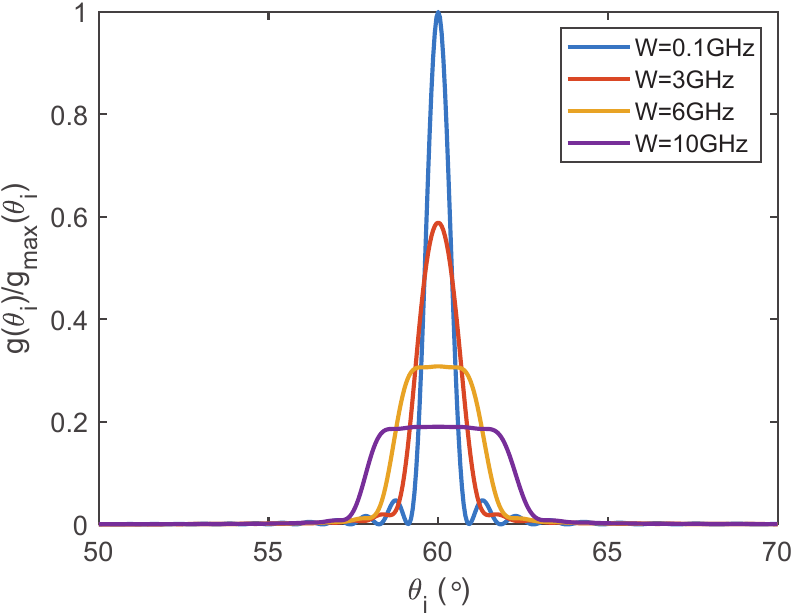}%
\label{fig_first_case}}
\hfil
\subfloat[]{\includegraphics[width=65mm]{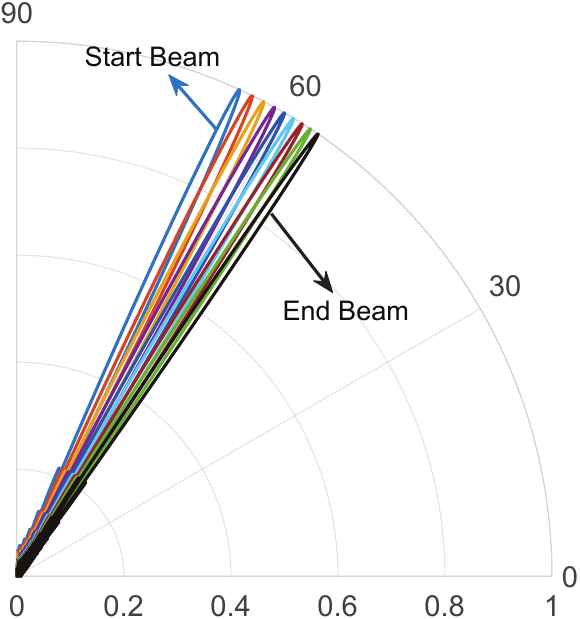}%
\label{fig_second_case}}
\caption{(a) An example of time-division beam sweeping, where the single target is located at $60^\circ$ and  $f_0 = 220$ GHz.
(b) Antenna array radiation pattern of BS under beam squint, where $f_0=220$ GHz, $W=22$ GHz and the central frequency subcarrier is focused at $60^\circ$.}
\label{fig_sim}
\end{figure*}

Based on (1) and (4),
the overall ISAC echo channel   on the $m$-th subcarrier of the $n_s$-th  symbol can be represented as
\begin{equation}
\begin{split}
\begin{aligned}
\label{deqn_ex1a}
\mathbf{H}_{n_s,m} = \sum_{k=1}^K \!\mathbf{H}_{k,n_s,m} + \sum_{i=1}^I\! \mathbf{H}_{i,n_s,m}'.
\end{aligned}
\end{split}
\end{equation}
Eq. (5) indicates that the echo channel of the ISAC system should be the sum of the background environment channel and the dynamic targets  channel.

\section{Sensing Oriented Beamforming  with Controllable Beam Squint}

Massive MIMO systems are usually combined with
 beamforming technology to compensate for the high propagation loss of  mmWave or Terahertz signals and enhance the directional communications and sensing performance\cite{8733134,7959180}.
However, directly
applying beamforming for  sensing requires
a long beam sweeping time as it did in existing ISAC   or the conventional  radar systems\cite{5262295,7390101}.
Meanwhile,
 it has been shown in \cite{wblsp} and \cite{9896734} that for a Terahertz communications system,
the  \emph{beam squint} effect would appear, in which the beams from different subcarriers would point to different directions, making the signals on some subcarriers deviate from the desired  direction.

\begin{figure*}[!t]
\centering
\subfloat[]{\includegraphics[width=70mm,height=60mm]{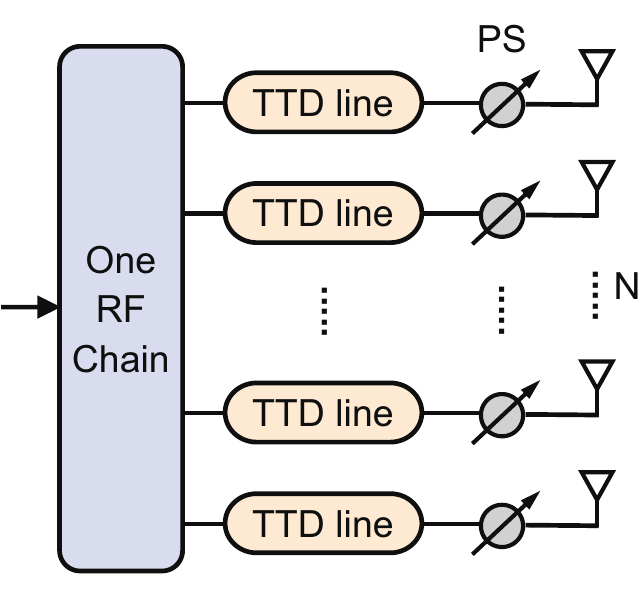}%
\label{fig_first_case}}
\hfil
\subfloat[]{\includegraphics[width=64mm]{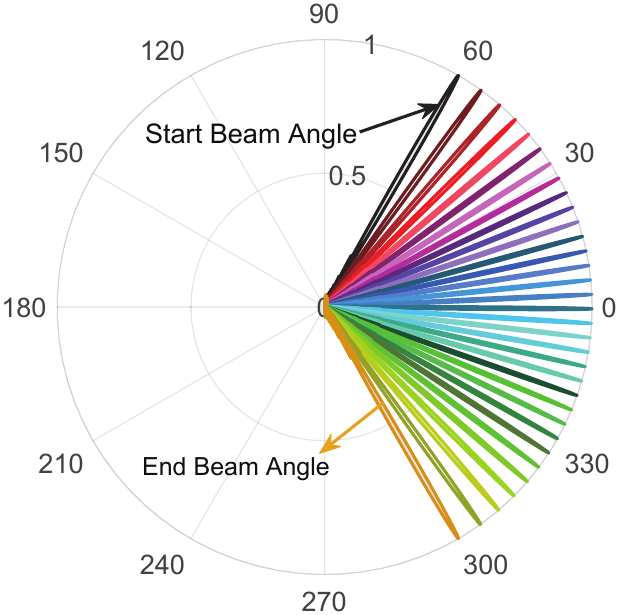}%
\label{fig_second_case}}
\caption{(a) TTDs-assisted BS antenna array links.
(b) An example of controllable beam squint.}
\label{fig_sim}
\end{figure*}

\subsection{The Influence of Beam Squint on Sensing}

Suppose
 the beamforming  vectors used  during transmitting and receiving signals are $\mathbf{w}_t$ and $\mathbf{w}_r$, respectively. Assume that the frequency-domain probing  data sent by the BS is $\mathbf{s}=\mathbf{1}\in \mathbb{C}^{(M+1)\times 1}$ and consider  the  noiseless case.
Then the received  echo signal on the $m$-th subcarrier of the $n_s$-th symbol is
\begin{equation}
\begin{split}
\begin{aligned}
\label{deqn_ex1a}
y_{n_s,m} = \mathbf{w}_r^H \mathbf{H}_{n_s,m} (\mathbf{w}_t^H)^T.
\end{aligned}
\end{split}
\end{equation}
Since the receiving and transmitting antennas are at the same location, the directions of departure and arrival associated with the target are the same. Hence
 $\mathbf{w}_r$ is generally the same as    $\mathbf{w}_t$\footnote{Even if the system works in half duplex mode\cite{9898900}, for the normal communications distance,
it can still be considered that the transmitting  and  receiving  array are at the same location.
Hence there is still  $\mathbf{w}_r=\mathbf{w}_t$
when the two arrays have the same configuration and are placed in parallel.}. In the traditional sensing,
 the BS searches the entire angle space via time domain   beam sweeping.
Suppose the BS uses  $L$ beam sweeps, where the $l$-th beam sweep points to angle $\theta_l$. Then the beamforming vector used for the $l$-th beam sweep is
\begin{equation}
\begin{split}
\begin{aligned}
\label{deqn_ex1a}
&\mathbf{w}_{t,l} = \mathbf{w}_{r,l} = \frac{1}{\sqrt{N}}\mathbf{a}(\theta_l)
\\&=\frac{1}{\sqrt{N}}[e^{j\frac{2\pi f_c}{c}(-\frac{N-1}{2})d\sin\theta_l},...,e^{j\frac{2\pi f_c}{c}(\frac{N-1}{2})d\sin\theta_l}]^T.
\end{aligned}
\end{split}
\end{equation}
Thus the total power of echo signals from all subcarriers  in the $l$-th beam sweep is
\begin{equation}
\begin{split}
\begin{aligned}
\label{deqn_ex1a}
g(\theta_l) &= \sum _{m=0}^{M}|y_{n_s,m,l}|=\sum _{m=0}^{M}|\mathbf{w}_{r,l}^H \mathbf{H}_{n_s,m} (\mathbf{w}_{t,l}^H)^T|\\&=\sum _{m=0}^{M}\left|\frac{1}{N}\mathbf{a}^H(\theta_l) \mathbf{H}_{n_s,m} (\mathbf{a}^H(\theta_l))^T\right|.
\end{aligned}
\end{split}
\end{equation}
Assume that the static clutter filtering has been performed on $y_{n_s,m,l}$.
Then the targets' angles can be estimated by
 finding the peak values of $g(\theta_l)$ after $L$ beam sweeps.

Fig.~3(a) shows an example of the curve $g(\theta_l)$ when a single target is fixed at $60^\circ$ and  $f_0=220$ GHz.
If the transmission bandwidth $W$ is small (i.e., the narrowband signal), then the angle of the target can be well distinguished from the peak  of $g(\theta_l)$.
However, if $W$ is large (i.e., the wideband signal), then the peak in  $g(\theta_l)$ gradually disappears, which deteriorates the resolution of angle estimation.
To explain this phenomenon,
we plot the antenna array radiation patterns on different subcarriers  in Fig.~3(b).
It is seen  that when the central frequency subcarrier points to  $60^\circ$ with
 $\mathbf{w}=\frac{1}{\sqrt{N}}\mathbf{a}(60^\circ)$, the beamforming directions of other subcarriers  gradually  squint to other angles.
Clearly, the beam squint effect leads to 
energy leakage for sensing, which then causes the flattening of the peak  in
Fig.~3(a).

\subsection{Utilizing  Beam Squint for Sensing}

To deal with this harmful  beam divergence problem,
the traditional solutions adopt the true time delay lines (TTDs) to re-concentrate
all subcarriers into one direction\cite{9896734}.
Nevertheless,  we can actually take  advantage of the beam squint effect to cover a wide area within one single OFDM symbol.
As shown in Fig.~4(a),
we assume that each antenna is configured with a phase shifter (PS) and a TTD.
Denote  the $n$-th phase shifter response as $e^{-j2\pi\phi_n}$.
The time domain response of the $n$-th TTD  is $\delta(t-t_n)$,
and its frequency domain response is $e^{-j2\pi \widetilde{f}t_n}$.
Then, the  array beamforming vector assisted by TTDs can be expressed as
\begin{equation}
\begin{split}
\begin{aligned}
\label{deqn_ex1a}
[\widetilde{\mathbf{w}}]_n=\frac{1}{\sqrt{N}}e^{-j2\pi\phi_n}e^{-j2\pi \widetilde{f}t_n},
\end{aligned}
\end{split}
\end{equation}
where
 $\phi_n$ denotes the phase shift  of the $n$-th PS,
$t_n$ denotes the time delay  of the $n$-th TTD,
and $\widetilde{f}$ is the baseband frequency.

The steering vector of any  angle direction $\theta$ on the $m$-th subcarrier is
 $\mathbf{a}_{m}(\theta) \in \mathbb{C}^{N\times 1}$, where
$[\mathbf{a}_{m}(\theta)]_n=e^{j\frac{2\pi f_m }{c}nd\sin\theta}.$
With TTDs, the array gain on the $m$-th subcarrier   is
\begin{equation}
\begin{split}
\begin{aligned}
\label{deqn_ex1a}
g_m&=\left|\widetilde{\mathbf{w}}^H\cdot \mathbf{a}_{m}(\theta)\right|\\&=\frac{1}{\sqrt{N}}
\left|
\sum _{n=-\frac{N-1}{2}}^{\frac{N-1}{2}}
e^{j2\pi\phi_n}e^{j2\pi \widetilde{f}_mt_n}e^{j\frac{2\pi f_m }{c}nd\sin\theta}\right|.
\end{aligned}
\end{split}
\end{equation}
It is seen from  (10) that
 $g_m$  reaches its maximum value when
\begin{equation}
\begin{split}
\begin{aligned}
\label{deqn_ex1a}
\phi_n + \widetilde{f}_mt_n + f_m\frac{ nd\sin\theta}{c}=0.
\end{aligned}
\end{split}
\end{equation}

By adjusting the values of  PSs,
we can let the beamforming at the $0$-th subcarrier $f_0$ point to a start angle $\theta_{start}$.
Specifically,
we set ${f}_m={f}_0$, $\widetilde{f}_m=\widetilde{f}_0=0$, $\theta=\theta_{start}$ in (11), and derive $\phi_{n}=-\frac{f_0nd\sin{\theta_{start}}}{c}$.
Then we set the phase shift   of the $n$-th PS as $\phi_{n}$.
On the other side, 
by adjusting the values of  TTDs,
we can let the beamforming at the $M$-th subcarrier $f_M$ point to an end angle $\theta_{end}$.
Specifically,
we set ${f}_m={f}_M$, $\widetilde{f}_m=\widetilde{f}_M=W$, $\theta=\theta_{end}$ in (11), and derive
$t_{n}=-\frac{\phi_n}{W}-\frac{(f_0+W)nd\sin{\theta_{end}}}{Wc}$. Then we set the time delay  of the $n$-th TTD line as $t_{n}$.
Afterwards, the beams from the $0$-th subcarrier to the $M$-th subcarrier  gradually squint from the start angle
$\theta_{start}$ to the end angle $\theta_{end}$ in a controlled way.
For subcarrier $f_m$, its beamforming angle $\theta_m$ still satisfies  (11)
and can be computed as follows.
We
substitute $\theta=\theta_m$,
$\phi_{n}=-\frac{f_0nd\sin{\theta_{start}}}{c}$,
$t_{n}=-\frac{\phi_n}{W}-\frac{(f_0+W)nd\sin{\theta_{end}}}{Wc}$  into (11), and then
 the beam squint angle ${\theta}_m$ should satisfy
\begin{equation}
\begin{split}
\begin{aligned}
\label{deqn_ex1a}
\sin{\theta}_m\!=\!\frac{(W\!-\!\widetilde{f}_{m})f_0}{Wf_m}\sin\theta_{start}\!+\!\frac{(W\!+\!f_0)\widetilde{f}_{m}}{Wf_m}\sin\theta_{end}.
\end{aligned}
\end{split}
\end{equation}

An example of controllable beam squint is shown in Fig~4(b), where
 $M=32$.
By adjusting   PSs and TTDs, we successfully  control the start angle of beam squint as $\theta_{start} = 60^\circ$ and the end angle as $\theta_{end} = -60^\circ$.
Then  all subcarriers gradually squint from $60^\circ$ to $-60^\circ$, which covers the entire  space.
When  $M$ is large enough,
all  subcarriers can cover the whole sensing area
 within  one single OFDM symbol.
When $M=2048$, the average interval between  adjacent subcarriers,
i.e., the resolution 
 is $\frac{120^\circ}{2048}=0.059^\circ$.
We  term  the above beamforming scheme  \emph{Beamforming Strategy with Controllable Beam Squint}, i.e., \emph{CBS-Beamforming} for short.

\section{Dynamic Target Sensing Scheme Based on CBS-Beamforming}

\begin{figure}[!t]
\centering
\includegraphics[width=80mm,height=60mm]{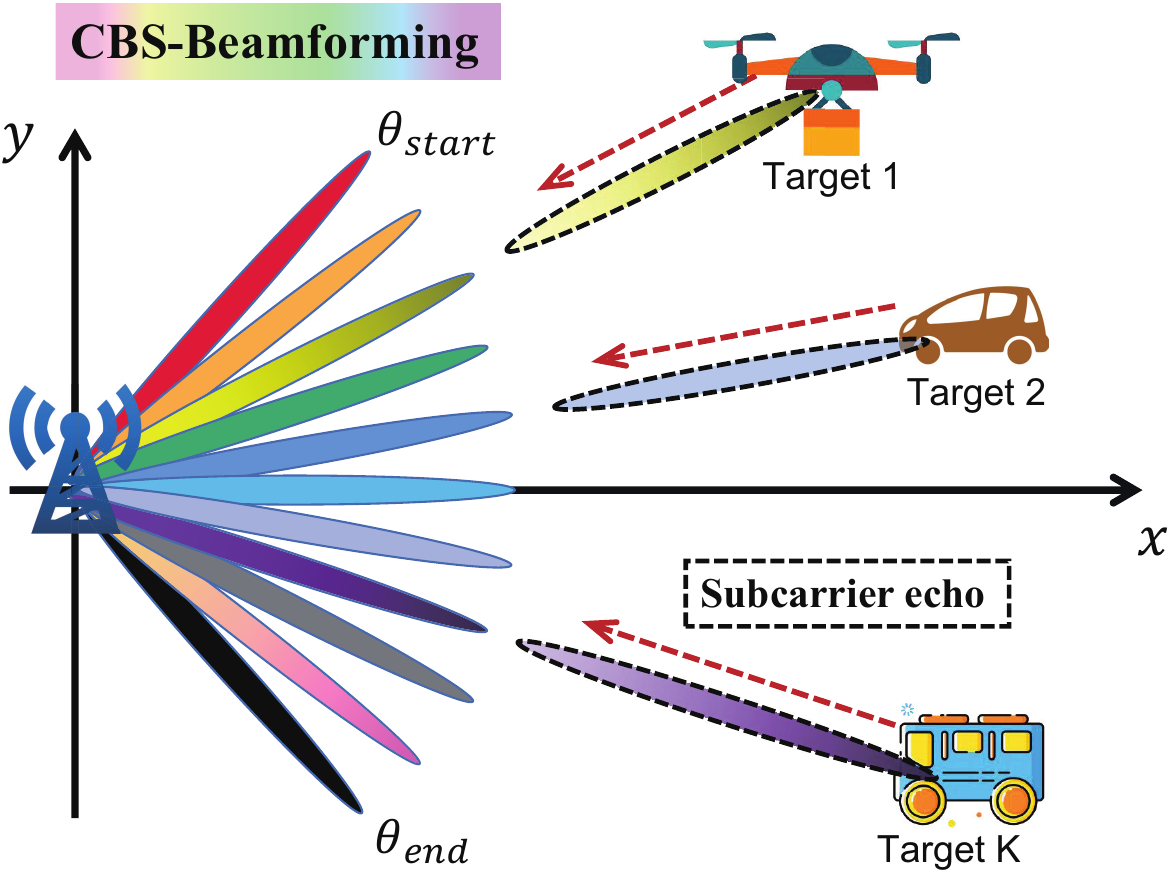}
\caption{Dynamic target sensing system diagram based on CBS-Beamforming.}
\label{fig_1}
\end{figure}

\subsection{You Only Listen Once (YOLO): A Fast Sensing Scheme Based on CBS-Beamforming}

Assume that the sensing range required by the BS is $[\theta_{start},\theta_{end}]$, and there are $K$  dynamic targets within the sensing range as shown in Fig.~5.
Without loss of generality, we denote their positions as $(r_1,\theta_1),(r_2,\theta_2),...,(r_K,\theta_K)$, where
$\theta_{start}>\theta_1>\theta_2>...>\theta_K>\theta_{end}$.
According to  (5), the echo channel matrix on the $m$-th subcarrier of the $n_s$-th symbol  is $\mathbf{H}_{n_s,m}$.

As illustrated in Fig. 5, the BS only needs  one-time beam sweeping
based on CBS-Beamforming to cover the whole sensing range, thanks to the dispersive effect of  beam squint.
With all $\mathbf{1}$ probing pilot,
 the received  echo signal on the $m$-th subcarrier of the $n_s$-th symbol at the BS is
\begin{equation}
\begin{split}
\begin{aligned}
\label{deqn_ex1a}
y_{n_s,m} &=\widetilde{\mathbf{w}}^H \mathbf{H}_{n_s,m}(\widetilde{\mathbf{w}}^H)^T
\\&=\sum_{k=1}^K \alpha_k e^{j\frac{4\pi f_0v_k}{c}n_sT_s}e^{-j\frac{4\pi f_mr_k}{c}}\widetilde{\mathbf{w}}^H \mathbf{a}_{k,m}\mathbf{a}_{k,m}^T\widetilde{\mathbf{w}}^* \\&  \quad + \sum_{i=1}^I \beta_i e^{-j\frac{4\pi f_mr'_i}{c}}\widetilde{\mathbf{w}}^H \mathbf{a}_{m}(\theta'_i)\mathbf{a}^T_{m}(\theta'_i)\widetilde{\mathbf{w}}^*.
\end{aligned}
\end{split}
\end{equation}
Then the echo signals on all subcarriers  and OFDM symbols can be combined into an echo signal matrix $\mathbf{Y} \in \mathbb{C}^{N_s\times (M+1)}$, and the $(n_s,m)$-th element in $\mathbf{Y}$ is $[\mathbf{Y}]_{n_s,m}=y_{n_s,m}$, where $n_s=0,1,...,N_s-1$ and $m=0,1,2,...,M$.

 For echo signal processing, some existing ISAC studies\cite{9898900,10050406,2023arXiv230109439M} only considered the sensing of static targets, and   overlooked the widespread static environment. In fact, it is not easy to distinguish static targets from the static environment. Another part of ISAC studies \cite{8918315,8114253,10048770,10011549,2023arXiv230512994H} only considered the sensing of dynamic targets, and did not consider the serious interference caused by static environmental clutter on dynamic target sensing.
 In practice, it is difficult to directly detect and estimate dynamic targets in  clutter backgrounds.
 Therefore, we must first
 filter out the dense echoes caused by static environments.

 Considering that each scattering unit of the static environment does not cause significant Doppler frequency shift within $N$ consecutive OFDM symbols, we average each column of the echo signal matrix $\mathbf{Y}$ and obtain the echo signal vector caused by the static environment as $\mathbf{y}_{static} \in \mathbb{C}^{1\times (M+1)}$, whose  $m$-th element is $[\mathbf{y}_{static}]_m = \frac{1}{N_s}\sum_{n_s=0}^{N_s-1}y_{n_s,m}$. 
Note that 
$\mathbf{y}_{static}$ can be used to reconstruct the echo signal matrix caused by the static environment as $\mathbf{Y}_{static}=[\mathbf{y}^T_{static},...,\mathbf{y}^T_{static}]^T\in \mathbb{C}^{N_s\times (M+1)}$.
Then the echo signal matrix caused by the dynamic targets can be extracted as
$\mathbf{Y}_{dynamic}\! =\! \mathbf{Y}\! -\! \mathbf{Y}_{static}$, and we represent the
 $(n_s,m)$-th element in $\mathbf{Y}_{dynamic}$ as 
\begin{equation}
\begin{split}
\begin{aligned}
\label{deqn_ex1a}
&\!\!\widetilde{y}_{n_s,m}=\sum_{k=1}^K \alpha_k e^{j\phi{(\!k,\!n_s,\!f_m)}}\widetilde{\mathbf{w}}^H \mathbf{a}_{k,m}\mathbf{a}_{k,m}^T\widetilde{\mathbf{w}}^* - \mathcal{I}'_{n_s,m}\\
&\!\!\!\!\!\!=\!\!\sum_{k=1}^{K}\!\!\left\{\frac{\!\!\alpha_k e^{j\phi{(\!k\!,\!n_s\!,\!f_m)}}}{N}\!\!
\left[\!\sum_{n=-\frac{N-1}{2}}^{\frac{N-1}{2}}\!\!\!\!\!\!\!e^{\!j2\pi\phi_{n}}
\!\!e^{j2\pi \widetilde{f}_{m}t_{n}}\!\!e^{\!j\!\frac{2\pi \!f_m}{c}\!n\!d\!\sin\!\theta_{k}}\!\!\right]^2\!\!\right\}\!-\!\! \mathcal{I}'_{n_s,m},
\end{aligned}
\end{split}
\end{equation}
where $e^{j\phi{(k,n_s,f_m)}}=e^{j\frac{4\pi f_0v_k}{c}n_sT_s}e^{-j\frac{4\pi f_mr_k}{c}}$, $\mathcal{I}'_{n_s,m}$ is the remaining term introduced by clutter filtering. It is easy to prove that 
$\lim\limits_{N_s \to \infty} \mathcal{I}'_{n_s,m} =0$. In  practical applications of radar systems, the term $\mathcal{I}'_{n_s,m}$ is usually  ignored\footnote{In actual systems, $N_s$ has to be truncated to a finite value, which leads to objective performance leakage in clutter suppression. But these performance leaks decrease as $N_s$ increases, hence they can and have to be tolerated by actual systems. This situation is similar to the theoretical research of massive MIMO system  that assumes an infinite number of antennas, but the actual MIMO  system has a limited number of antennas.}. Then
we bring $\phi_{n}=-\frac{f_0nd\sin\theta_{start}}{c}$ and $t_{n}=-\frac{\phi_{n}}{W}-\frac{(f_0+W)nd\sin\theta_{end}}{Wc}$ into (14) and obtain
\begin{equation}
\begin{split}
\begin{aligned}
\label{deqn_ex1a}
\widetilde{y}_{n_s,m}\approx \!\!
\frac{1}{N} \sum_{k=1}^{K}\left\{\alpha_k e^{j\phi{(k,n_s,f_m)}}\!\!\!
\left[\frac{\sin(\frac{\beta(f_m,\theta_k)}{2}N)}{\sin(\frac{\beta(f_m,\theta_k)}{2})}\right]^2\right\},
\end{aligned}
\end{split}
\end{equation}
where
\begin{equation}
\begin{split}
\begin{aligned}
\label{deqn_ex1a}
&\!\!\!\!\beta(f_m,\theta_k)=\frac{2\pi d}{Wc}\times 
\\&\!\!\!\![(W\!\!-\!\!\widetilde{f}_{m})\!f_0\!\sin\!\theta_{start}\!+\!(W\!\!+\!\!f_0)\!\widetilde{f}_{m}\sin\!\theta_{end}\!-\!W\!f_m\sin\theta_{k}].
\end{aligned}
\end{split}
\end{equation}
The detailed derivation of  (15) can be found  in Appendix A.

By taking the modulus  of $\widetilde{y}_{n_s,m}$,
we can obtain the corresponding signal power as  $g_{n_s,m}=\left|\widetilde{y}_{n_s,m}\right|$.
Then the  echo signal powers  of all subcarriers and symbols   can be pieced into a matrix $\mathbf{G} \in \mathbb{C}^{N_s\times (M+1)}$, which  is called the \emph{power spectrum matrix} of the echoes and satisfies $[\mathbf{G}]_{n_s,m}=g_{n_s,m}$.
In addition, for the $n_s$-th OFDM symbol, the
signal power  of all subcarriers received during this symbol time  can be pieced into a  \emph{power spectrum vector} $\mathbf{g}_{n_s}=[g_{n_s,0},g_{n_s,1},...,g_{n_s,M}]^T \in \mathbb{C}^{(M+1)\times 1}$. Thus we have   
\begin{equation}
\begin{split}
\begin{aligned}
\label{deqn_ex1a}
\mathbf{G} = [\mathbf{g}_{0}, \mathbf{g}_{1},...,\mathbf{g}_{N_s-1}]^T.
\end{aligned}
\end{split}
\end{equation}

When $n_s$ is fixed,  i.e., within the same OFDM symbol,
it can be seen from (15) and (16) that  $g_{n_s,m}$  changes with the frequency $f_m$, and the  powers of different subcarriers are different.
In order to analyze the impacts from different targets,
we designate a specific target as  the $k^*$-th target, whose parameters are  denoted as $r_{k^*}$, $\theta_{k^*}$ and $v_{k^*}$.
For given $\theta_{k^*}$ in  (16),
 $\beta(f_m,\theta_{k^*})$ is a linear function of $f_m$, where $f_m=f_0+\widetilde{f}_{m}$ and  $f_0\leq f_m\leq f_M =  f_0+W$.
Since  $\theta_{k^*}$ belongs to $[\theta_{start},\theta_{end}]$,
the product of the linear function's terminal values is
$\beta(f_0,\theta_{k^*})\cdot\beta(f_M,\theta_{k^*})=\frac{2\pi df_0}{c}(\sin\theta_{start}-\sin\theta_{k^*})\cdot \frac{2\pi df_M}{c}(\sin\theta_{end}-\sin\theta_{k^*})<0$. Hence  there must exist an appropriate $f_m$ that makes
$\beta(f_m,\theta_{k^*})=0$ when the number of subcarriers $M+1$ is sufficiently large.
Denote this frequency as $f^{d,k^*}_{m}$ and thus the constraint  between $f^{d,k^*}_{m}$ and $\theta_{k^*}$ is  $\beta(f^{d,k^*}_{m},\theta_{k^*})=0$.
Let us then consider $2\ddot{M}+1$ subcarriers around the  subcarrier $f^{d,k^*}_{m}$, whose  frequencies can be denoted as $f^{d,k^*}_{m,-\ddot M},...,f^{d,k^*}_{m,0},...,f^{d,k^*}_{m,\ddot M}$ with  $f^{d,k^*}_{m,\ddot m}=f^{d,k^*}_{m}+\ddot m
\frac{W}{M}$, $\ddot m=-\ddot M,...,\ddot M$ and $2\ddot{M}+1 \ll M+1$.
We record the beam squint angle corresponding to the subcarrier $f^{d,k^*}_{m,\ddot m}$ as
$\theta^{d,k^*}_{m,\ddot m}$.
It can be seen from (14) that $f^{d,k^*}_{m,\ddot m}$ and $\theta^{d,k^*}_{m,\ddot m}$ satisfy
$\beta(f^{d,k^*}_{m,\ddot m},\theta^{d,k^*}_{m,\ddot m})=0$, i.e., $(W-\widetilde f^{d,k^*}_{m,\ddot m})f_0\sin\theta_{start}+(W+f_0)\widetilde f^{d,k^*}_{m,\ddot m}\sin\theta_{end}=Wf^{d,k^*}_{m,\ddot m}\sin{\theta}^{d,k^*}_{m,\ddot m}$.
Then
we substitute  $f_m=f^{d,k^*}_{m,\ddot m}$ and $\beta(f^{d,k^*}_{m,\ddot m},\theta^{d,k^*}_{m,\ddot m})=0$ into (16) and obtain
\begin{equation}
\begin{split}
\begin{aligned}
\label{deqn_ex1a}
\beta(f^{d,k^*}_{m,\ddot m},\theta_{k})&=\frac{2\pi d}{Wc}[Wf^{d,k^*}_{m,\ddot m}\sin\theta^{d,k^*}_{m,\ddot m}\!-\!Wf^{d,k^*}_{m,\ddot m}\sin\theta_{k}]\\&
=\frac{2\pi d f^{d,k^*}_{m,\ddot m}}{c}(\sin\theta^{d,k^*}_{m,\ddot m}-\sin\theta_k).
\end{aligned}
\end{split}
\end{equation}

\begin{figure*}[!t]
\centering
\subfloat[]{\includegraphics[width=70mm]{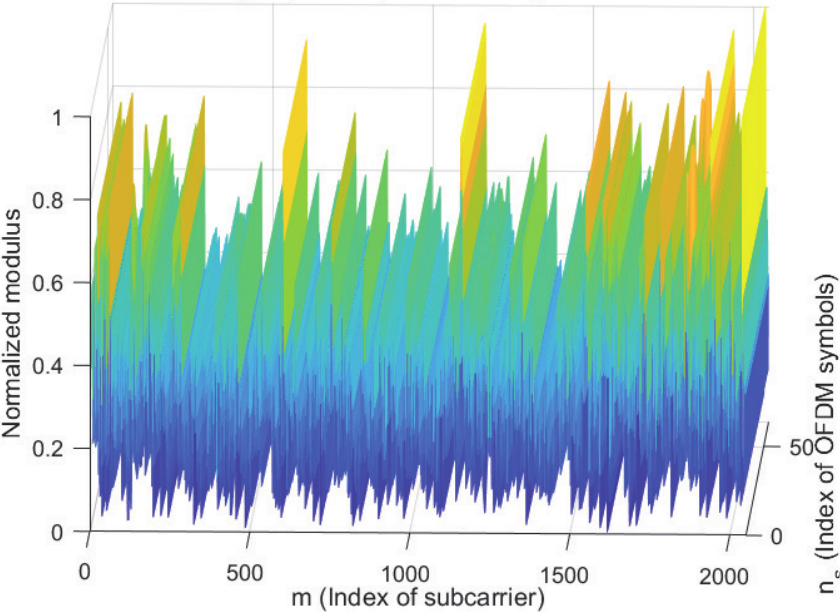}%
\label{fig_first_case}}
\hfil
\subfloat[]{\includegraphics[width=70mm]{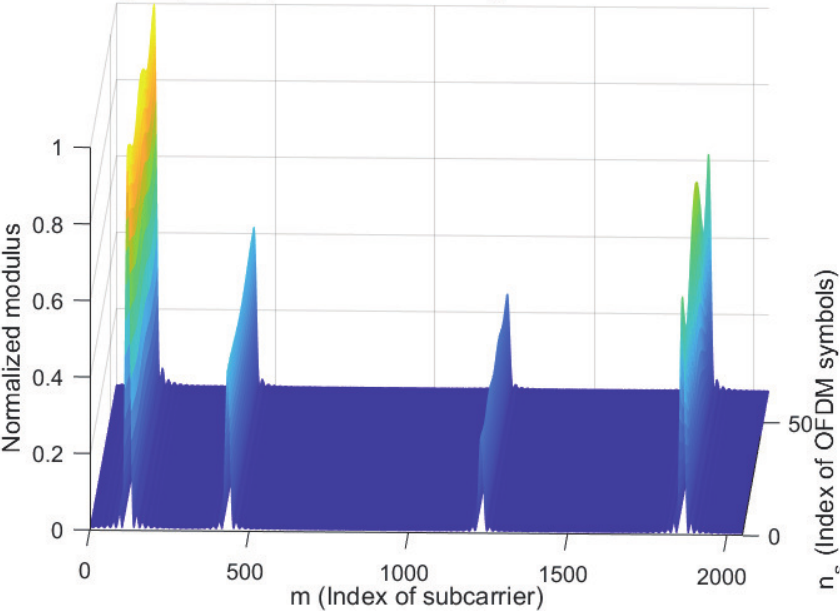}%
\label{fig_first_case}}
\hfil
\subfloat[]{\includegraphics[width=70mm]{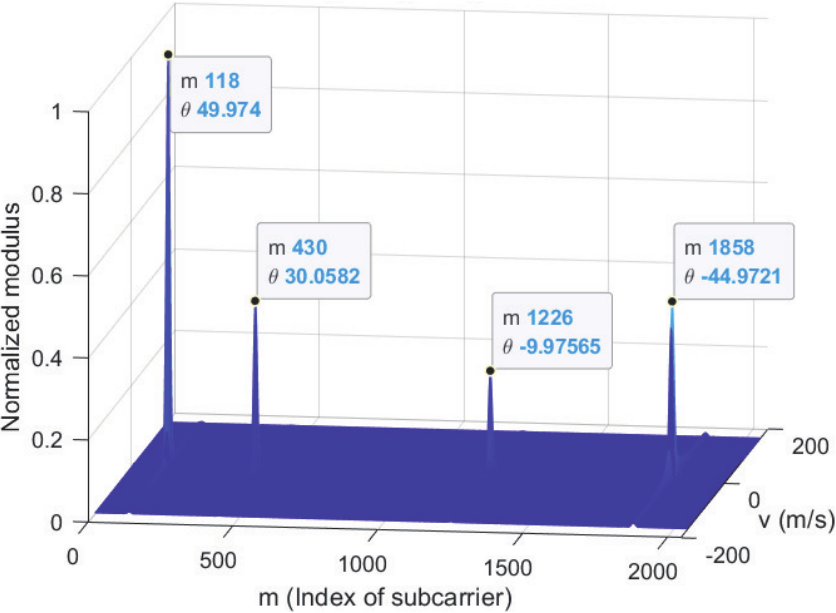}%
\label{fig_first_case}}
\hfil
\subfloat[]{\includegraphics[width=70mm]{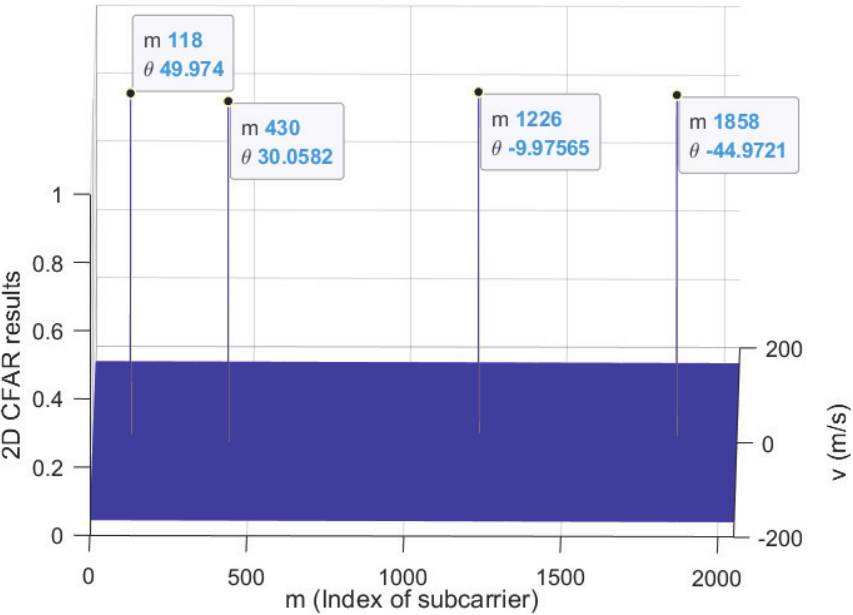}%
\label{fig_second_case}}
\caption{(a) Power spectrum of  original echo signals. (b) Signal power spectrum after filtering out static environmental clutter.
(c) The angle-Doppler spectrum after velocity FFT.
(d) The detection results of 2D-CFAR on angle-Doppler spectrum.
The four dynamic targets are set as $(50m,50^\circ,10m/s)$, $(120m,30^\circ,-5m/s)$, $(200m,-10^\circ,15m/s)$ and $(80m,-45^\circ,8m/s)$. $[\theta_{start},\theta_{end}]=[60^\circ,-60^\circ]$,
$N=128$ and $M+1=2048$.}
\label{fig_sim}
\end{figure*}

Substituting  (18) into (15),
 the  echo signal of the subcarrier $f^{d,k^*}_{m,\ddot m}$ in the $n_s$-th OFDM symbol  is
\begin{equation}
\begin{split}
\begin{aligned}
\label{deqn_ex1a}
&\widetilde{y}^{d,k^*}_{n_s,m,\ddot m}\!\!
=\frac{1}{N} \sum_{k=1}^{K}\alpha_k e^{j\phi{(k,n_s,f^{d,k^*}_{m,\ddot m})}}
\left[\frac{\sin(\frac{\beta(f^{d,k^*}_{m,\ddot m},\theta_{k})}{2}N)}{\sin(\frac{\beta(f^{d,k^*}_{m,\ddot m},\theta_{k})}{2})}\right]^2\\
&\!\!\!\!=\!\!\frac{1}{N} \!\!\sum_{k=1}^{K}\!\!\alpha_k e^{j\phi{(k,n_s,f^{d,k^*}_{m,\ddot m})}}\!\!
\left\{ \!\!
\frac{\sin\left[\frac{\pi d f^{d,k^*}_{m,\ddot m}}{c}(\sin\theta^{d,k^*}_{m,\ddot m}\!\!-\!\sin\theta_k)N\right]}
{\sin\left[\frac{\pi d f^{d,k^*}_{m,\ddot m}}{c}(\sin\theta^{d,k^*}_{m,\ddot m}-\sin\theta_k)\right]}
\right\}^2\\
&\!\!\!\!=\!\!\frac{\!\alpha_{k^*} \! e^{\!j\!\phi{\!(k^*\!,\!n_s\!,\!f^{d,k^*}_{m,\ddot m}\!)}}}{N}\!\!\!
\left\{\!\!
\frac{\sin\!\!\left[\!\!\frac{\pi d f^{d,k^*}_{m,\ddot m}}{c}\!(\sin\theta^{d,k^*}_{m,\ddot m}\!\!-\!\sin\!\theta_{k^*}\!\!)N\!\!\right]}
{\sin\!\!\left[\!\!\frac{\pi d f^{d,k^*}_{m,\ddot m}}{c}(\sin\theta^{d,k^*}_{m,\ddot m}\!\!-\!\sin\theta_{k^*})\!\right]}
\!\!\!\right\}^2\!\!\!\!\!
+\!
\mathcal{\ddot I}(n_s\!,\!f^{d,k^*}_{m,\ddot m}),
\end{aligned}
\end{split}
\end{equation}
where $e^{j\phi{(k,n_s,f^{d,k^*}_{m,\ddot m})}}= e^{j\frac{4\pi f_0v_k}{c}n_sT_s} e^{-j\frac{4\pi f^{d,k^*}_{m,\ddot m}r_k}{c}}$, 
$e^{j\phi{(k^*,n_s,f^{d,k^*}_{m,\ddot m})}} = e^{j\frac{4\pi f_0v_{k^*}}{c}n_sT_s} e^{-j\frac{4\pi f^{d,k^*}_{m,\ddot m}r_{k^*}}{c}}$, and 
\begin{equation}
\begin{split}
\begin{aligned}
\label{deqn_ex1a}
&\mathcal{\ddot I}(n_s,f^{d,k^*}_{m,\ddot m})\\&=\!\! \frac{1}{N}\!\!\!\!
\sum_{k=1,k\neq k^*}^{K}\!\!\!\!\!\!\!\!\!\alpha_k e^{j\phi{(k,n_s,f^{d,k^*}_{m,\ddot m})}}\!\!
\left\{\!
\frac{\sin\!\!\left[\!\!\frac{\pi d f^{d,k^*}_{m,\ddot m}}{c}(\sin\theta^{d,k^*}_{m,\ddot m}\!\!-\!\sin\theta_k)N\right]}
{\sin\left[\frac{\pi d f^{d,k^*}_{m,\ddot m}}{c}(\sin\theta^{d,k^*}_{m,\ddot m}\!\!-\!\sin\theta_k)\right]}
\!\right\}^2\!\!
\end{aligned}
\end{split}
\end{equation}
 is the interference term of the other $K-1$ targets to the $k^*$-th target.
For massive MIMO systems with sufficiently large $N$, we have  
 $\mathcal{\ddot I}(n_s,f^{d,k^*}_{m,\ddot m})\approx 0$, as long as $\theta^{d,k^*}_{m,\ddot m}\neq\theta_k$ in $\mathcal{\ddot I}(n_s,f^{d,k^*}_{m,\ddot m})$\cite{7524027}.
In addition, noticing   $\theta^{d,k^*}_{m,0}=\theta_{k^*}$ and $f^{d,k^*}_{m}=f^{d,k^*}_{m,0}$,   we can obtain from (19) that
\begin{align}
\widetilde{y}^{d,k^*}_{n_s,m}&=\widetilde{y}^{d,k^*}_{n_s,m,0}\!\approx\! \alpha_{k^*} N e^{\!j\!\frac{4\pi f_0v_{k^*}}{c}n_sT_s} e^{-j\frac{4\pi f^{d,k^*}_{m}r_{k^*}}{c}}\label{1},\\
g^{d,k^*}_{n_s,m}&= \left|\widetilde{y}^{d,k^*}_{n_s,m}\right| \approx \alpha_{k^*} N \label{2},\\
\varphi^{d,k^*}_{n_s,m,\ddot m}&=\! \arg\{\!\widetilde{y}^{d,k^*}_{n_s,m,\ddot m}\}\! \approx \!
\frac{4\pi \!f_0v_{k^*}n_s\!T_s}{c} \!-\!\frac{4\pi \!f^{d,k^*}_{m,\ddot m}r_{k^*}}{c} \label{3},
\end{align}
where $\widetilde{y}^{d,k^*}_{n_s,m}$ and  $g^{d,k^*}_{n_s,m}$   represent the
complex echo signal and  the echo signal power of the subcarrier $f^{d,k^*}_{m}$ in the $n_s$-th OFDM symbol  respectively, while  $\varphi^{d,k^*}_{n_s,m,\ddot m}$ represents the theoretical phase
of the subcarrier $f^{d,k^*}_{m,\ddot m}$ in the $n_s$-th  symbol.

It is clear from (22)  that the power of the subcarrier $f^{d,k^*}_{m}$ is a peak value in the  power spectrum vector $\mathbf{g}_{n_s}$, and 
we call this subcarrier  a \emph{peak power subcarrier}.
Under noiseless conditions, the frequency of the peak power subcarriers corresponding to the same target detected by the BS from $\mathbf{g}_{n_s}$ within each OFDM symbol  ($n_s=0,1,2,...,N_s-1$) should be the same.
Therefore, in  noisy environments, the BS should detect a unique peak power subcarrier  corresponding to a certain target from the dynamic echo matrix $\mathbf{Y}_{dynamic}$.
For this effect,  by performing an $N_s$-point Fast Fourier Transform (FFT), i.e., a  velocity FFT, on each column of $\mathbf{Y}_{dynamic}$, and by moving the zero frequency component of the transformed spectrum to the center of the array, the Angle-Doppler spectrum of the echo signal  can be obtained as
${\rm \mathbf{Y}}_{AD} = {\rm FFTshift}\{{\rm FFT}\{\mathbf{Y}_{dynamic},N,1\},N,1\}$. The corresponding Angle-Doppler power spectrum is represented as ${\rm \mathbf{G}}_{AD}$, which satisfies $[{\rm \mathbf{G}}_{AD}]_{n_s,m}=|[{\rm \mathbf{Y}}_{AD}]_{n_s,m}|$. Then by performing two-dimensional  Constant False Alarm Rate (CFAR) detection on ${\rm \mathbf{G}}_{AD}$, $K$ peak values can be detected corresponding to $K$ dynamic targets one by one, and the frequency  of the unique peak power subcarrier corresponding to the $k^*$-th dynamic target can be obtained as $f^{d,k^*}_{m}$ from the 2D detection results.
Therefore, when the BS receives the echo signal and detects
the subcarrier with the peak power,
the angle estimation result of the $k^*$-th dynamic target can be obtained by solving $\beta(f^{d,k^*}_{m},\theta_{k^*})=0$  as
\begin{equation}
\begin{split}
\begin{aligned}
\label{deqn_ex1a}
\!\!\!\!\!\!\!\sin{\hat{\theta}}_{k^*}\!\!=\!\!\frac{(W\!\!-\!\widetilde{f}^{d,k^*}_{m})f_0}{Wf^{d,k^*}_{m}}\!\sin\theta_{start}\!+\!\frac{(W\!\!+\!f_0)\widetilde{f}^{d,k^*}_{m}}{Wf^{d,k^*}_{m}}\!\sin\theta_{end}.
\end{aligned}
\end{split}
\end{equation}
We record the frequencies of the peak power subcarriers corresponding to $K$ dynamic targets as
$\{f^{d,1^*}_{m},f^{d,2^*}_{m},...,f^{d,K^*}_{m}\}$.
Then the angle estimation  of all the $K$  dynamic targets $\{\hat{\theta}_{1},\hat{\theta}_{2},...,\hat{\theta}_{K}\}$
can  subsequently be determined through (24).

Fig.~6 shows an example of  static  clutter filtering, Angle-Doppler spectrum estimation, 2D-CFAR detection, and dynamic target angle estimation processes mentioned above,  where four dynamic targets are set as $(50m,50^\circ,10m/s)$, $(120m,30^\circ,-5m/s)$, $(200m,-10^\circ,15m/s)$ and $(80m,-45^\circ,8m/s)$.
It can be seen from Fig.~6(a) that the original echo signal carries a large amount of clutter caused by the static environment. After performing mean phasor cancellation, Fig.~6(b) retains almost only those echoes caused by the dynamic targets, and the angle information of the dynamic target is directly related to the frequency of the peak power subcarrier.
After performing a velocity FFT on the complex signal in Fig.~6(b), four peaks corresponding to four dynamic targets can be clearly observed in the Angle-Doppler spectrum shown in Fig.~6(c), which can also be observed in the 2D-CFAR detection results shown in Fig.~6(d).
 According to the peak power subcarriers and   (24), the angle estimates corresponding to  the dynamic targets  can be obtained as $49.974^\circ$, $30.0582^\circ$, $-9.97565^\circ$ and $-44.9721^\circ$.

Next, to estimate the distance and velocity of   dynamic target,
let us name the $2\ddot{M}+1$ subcarriers around the peak power subcarrier $f^{d,k^*}_{m}$   \emph{peak sidelobe subcarriers},
mark the subcarrier with frequency $f^{d,k^*}_{m}$  as $m_{k^*}$,
and regard $2\ddot{M}+1$ as  the \emph{sidelobe window width}.
Then the echo signal matrix of the $2\ddot{M}+1$ peak sidelobe subcarriers corresponding to the $k^*$-th target within $N_s$ OFDM symbols can be extracted from $\mathbf{Y}_{dynamic}$ and  are represented as
$\mathbf{Y}_{RD,k^*} = \mathbf{Y}_{dynamic}[:,m_{k^*}-\ddot{M}:m_{k^*}+\ddot{M}] \in \mathbb{C}^{N_s\times (2\ddot{M}+1)}$.
It is seen from (23)   that the echo signals on these sidelobe subcarriers  carry the distance and velocity information of the $k^*$-th dynamic target.

\begin{figure*}[!t]
\centering
\subfloat[]{\includegraphics[width=70mm]{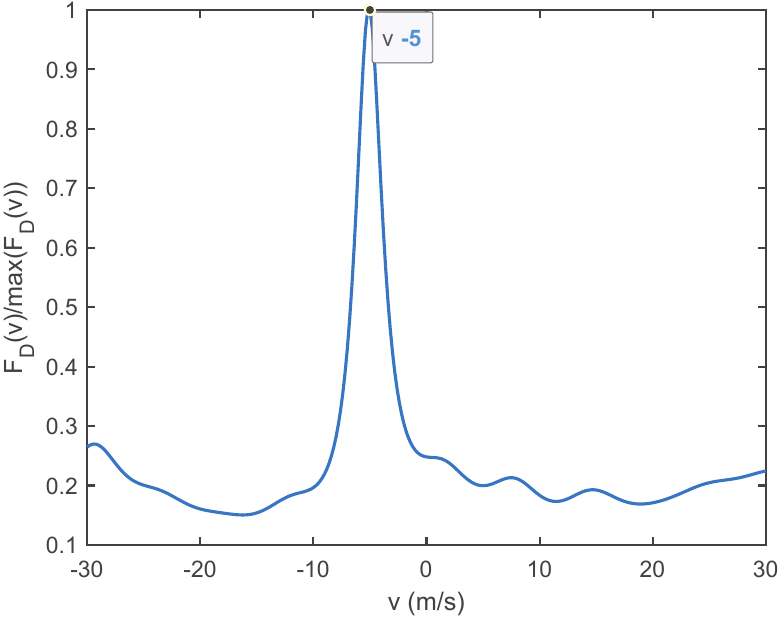}%
\label{fig_first_case}}
\hfil
\subfloat[]{\includegraphics[width=70mm]{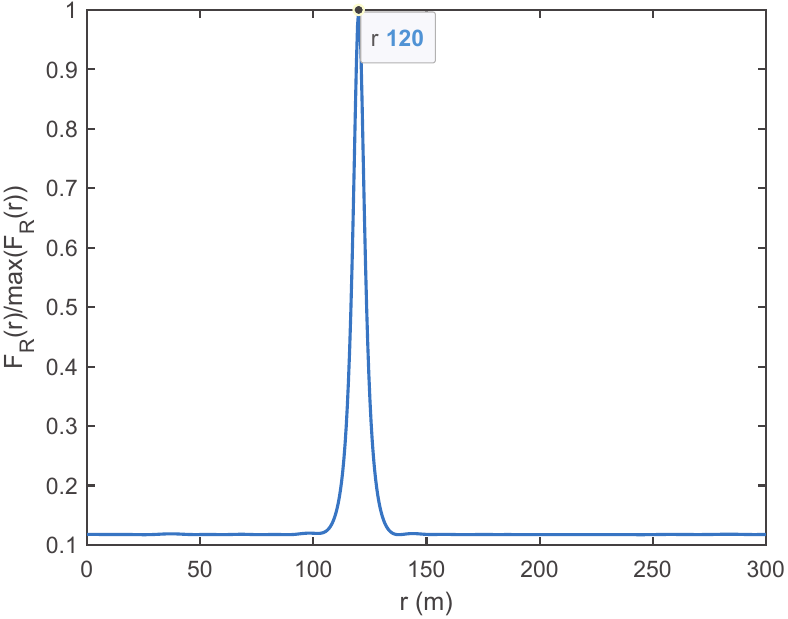}%
\label{fig_second_case}}
\caption{(a) An example of estimating dynamic target velocity based on the MUSIC algorithm.
(b) An example of estimating dynamic target distance based on the MUSIC algorithm.
The  target is set as $(120m,30^\circ,-5m/s)$ and SNR $= 0$ dB.}
\label{fig_sim}
\end{figure*}

Specifically, we  normalize each element in $\mathbf{Y}_{RD,k^*}$ to obtain a new sidelobe subcarrier echo signal matrix $\mathbf{\bar{Y}}_{RD,k^*}$, which satisfies $[\mathbf{\bar{Y}}_{RD,k^*}]_{i,j}=\frac{[\mathbf{Y}_{RD,k^*}]_{i,j}}{|[\mathbf{Y}_{RD,k^*}]_{i,j}|}$.
According to  (19) and (23), $\mathbf{\bar{Y}}_{RD,k^*}$ can be represented in detail as
\begin{equation}
\begin{split}
\begin{aligned}
\label{deqn_ex1a}
&\!\!\!\!\mathbf{\bar{Y}}_{RD,k^*}\!\! =\! e^{-j\frac{4\pi f^{d,k^*}_{m,-\ddot M}r_{k^*}}{c}}\times\\&
\begin{bmatrix}
	e^{\!j\frac{\!4\!\pi \!f_0v_{k^*}T_s}{c}0} e^{-j\frac{4\pi r_{k^*} \Delta f }{c}0}   & \!\!\!\!\cdots\!\!\!\!   & e^{\!j\!\frac{\!4\!\pi \!f_0\!v_{k^*}\!T_s}{c}0} e^{-j\frac{4\pi r_{k^*}\Delta f}{c}(2\ddot M)}   \\
	\vdots   & \ddots   & \vdots  \\
	e^{\!j\!\frac{4\!\pi \!f_0\!v_{k^*}\!T_s}{c}\!(\!N_s\!-\!1\!)} \!e^{\!-\!j\!\frac{4\pi r_{k^*}\!\Delta \!f}{c}\!0}  & \!\!\!\!\cdots\!\!\!\!   & \!\!e^{\!j\!\frac{\!4\!\pi \!f_0\!v_{k^*}\!T_s}{c}\!(\!N_s\!-\!1\!)} \!e^{\!-\!j\frac{\!4\pi r_{k^*}\!\Delta \!f}{c}(2\ddot M)} \\
\end{bmatrix}.
\end{aligned}
\end{split}
\end{equation}
We  define the Doppler steering vector in the $N_s$ dimensions and the distance steering vector in the $2\ddot M+1$ dimensions as
$\mathbf{k}_{D}(v_{k^*})=[1,e^{j\frac{4\pi f_0v_{k^*}T_s}{c}},...,e^{j\frac{4\pi f_0v_{k^*}T_s}{c}(N_s-1)}]^T \in \mathbb{C}^{N_s\times 1} $ and $\mathbf{k}_{R}(r_{k^*})=[1, e^{-j\frac{4\pi r_{k^*} \Delta f }{c}},...,e^{-j\frac{4\pi r_{k^*} \Delta f }{c}(2\ddot M)}    ]^T \in \mathbb{C}^{(2\ddot{M}+1) \times 1}$, respectively. Then $\mathbf{\bar{Y}}_{RD,k^*}$ and its  transpose matrix  can be further represented as
\begin{equation}
\begin{split}
\begin{aligned}
\label{deqn_ex1a}
\mathbf{\bar{Y}}_{RD,k^*} = e^{-j\frac{4\pi f^{d,k^*}_{m,-\ddot M}r_{k^*}}{c}}\mathbf{k}_{D}(v_{k^*}) \mathbf{k}^T_{R}(r_{k^*}),
\end{aligned}
\end{split}
\end{equation}
\begin{equation}
\begin{split}
\begin{aligned}
\label{deqn_ex1a}
\mathbf{\bar{Y}}^T_{RD,k^*} = e^{-j\frac{4\pi f^{d,k^*}_{m,-\ddot M}r_{k^*}}{c}}\mathbf{k}_{R}(r_{k^*}) \mathbf{k}^T_{D}(v_{k^*}).
\end{aligned}
\end{split}
\end{equation}
From  (26) and (27), it can be seen that $\mathbf{\bar{Y}}_{RD,k^*}$ and $\mathbf{\bar{Y}}^T_{RD,k^*}$ are the generalized array signal forms for  Doppler steering array and range steering array.
Hence  the velocity and distance of the $k^*$-th dynamic target can be estimated from the autocorrelation matrices of $\mathbf{\bar{Y}}_{RD,k^*}$ and $\mathbf{\bar{Y}}^T_{RD,k^*}$,  as
\begin{equation}
\begin{split}
\begin{aligned}
\label{deqn_ex1a}
\mathbf{R}^X_{D,k^*} =\frac{1}{2\ddot M+1}\mathbf{\bar{Y}}_{RD,k^*}  (\mathbf{\bar{Y}}_{RD,k^*} )^H,
\end{aligned}
\end{split}
\end{equation}
\begin{equation}
\begin{split}
\begin{aligned}
\label{deqn_ex1a}
\mathbf{R}^X_{R,k^*} =\frac{1}{N_s}\mathbf{\bar{Y}}^T_{RD,k^*}  (\mathbf{\bar{Y}}^T_{RD,k^*} )^H.
\end{aligned}
\end{split}
\end{equation}
We here adopt the reliable multiple signal classification (MUSIC) algorithm to solve the generalized array signal estimation problem.
We decompose the eigenvalues of $\mathbf{R}^X_{D,k^*}$ and $\mathbf{R}^X_{R,k^*}$ to obtain the diagonal matrix with eigenvalues ranging from large to small ($\mathbf{\Sigma}_{D,k^*}$ and $\mathbf{\Sigma}_{R,k^*}$) and the corresponding eigenvector matrix ($\mathbf{U}_{D,k^*}$ and $\mathbf{U}_{R,k^*}$). That is
\begin{equation}
\begin{split}
\begin{aligned}
\label{deqn_ex1a}
[\mathbf{U}_{D,k^*}, \mathbf{\Sigma}_{D,k^*}]={\rm eig}(\mathbf{R}^X_{D,k^*}),
\end{aligned}
\end{split}
\end{equation}
\begin{equation}
\begin{split}
\begin{aligned}
\label{deqn_ex1a}
[\mathbf{U}_{R,k^*}, \mathbf{\Sigma}_{R,k^*}]={\rm eig}(\mathbf{R}^X_{R,k^*}).
\end{aligned}
\end{split}
\end{equation}

Then  the Minimum Description Length (MDL) criterion is utilized to estimate the number of dynamic targets from $\mathbf{\Sigma}_{D,k^*}$ and $\mathbf{\Sigma}_{R,k^*}$  as $N_{k^*,D}^{MDL}$ and $N_{k^*,R}^{MDL}$ respectively, and  we define $N_{k^*}^{MDL}={\rm min}\{N_{k^*,D}^{MDL}, N_{k^*,R}^{MDL}\}$ as the number of dynamic targets to be estimated from $\mathbf{\bar{Y}}_{RD,k^*}$.
Under normal circumstances, there is always $N_{k^*,D}^{MDL} = N_{k^*,R}^{MDL}$.
Due to the assumption that the angles of each dynamic target are different from each other, there will be $N_{k^*}^{MDL}=1$.
Therefore, the noise space related to the velocity array can be represented as $\mathbf{U}^N_{D,k^*} = \mathbf{U}_{D,k^*}[:,N_{k^*}^{MDL}+1:N_s]$, and the noise space related to the distance array can be represented as $\mathbf{U}^N_{R,k^*} = \mathbf{U}_{R,k^*}[:,N_{k^*}^{MDL}+1:2\ddot M+1]$.
Then the  Doppler spectral function with search velocity $v$  and the  distance spectral function with search distance $r$ for the $k^*$-th dynamic target are
\begin{equation}
\begin{split}
\begin{aligned}
\label{deqn_ex1a}
F_{D,k^*}(v) = \frac{1}{\mathbf{k}^H_{D}(v)\mathbf{U}^N_{D,k^*}(\mathbf{U}^N_{D,k^*})^H\mathbf{k}_{D}(v)},
\end{aligned}
\end{split}
\end{equation}
\begin{equation}
\begin{split}
\begin{aligned}
\label{deqn_ex1a}
F_{R,k^*}(r) = \frac{1}{\mathbf{k}^H_{R}(r)\mathbf{U}^N_{R,k^*}(\mathbf{U}^N_{R,k^*})^H\mathbf{k}_{R}(r)}.
\end{aligned}
\end{split}
\end{equation}

Then  the velocity and distance estimation results of the $k^*$-th target can be obtained by searching the maximum values of  $F_{D,k^*}(v)$ and $F_{R,k^*}(r)$ , i.e,
\begin{equation}
\begin{split}
\begin{aligned}
\label{deqn_ex1a}
\!\!\!\!\hat{v}_{k^*}\!=\!\mathop{\arg\max}\limits_{v}F_{D,k^*}(v),
\end{aligned}
\end{split}
\end{equation}
\begin{equation}
\begin{split}
\begin{aligned}
\label{deqn_ex1a}
\!\!\!\!\hat{r}_{k^*}\!=\!\mathop{\arg\max}\limits_{r}F_{R,k^*}(r).
\end{aligned}
\end{split}
\end{equation}

Since the CBS-Beamforming can cover the entire sensing area in one beam sweep, the proposed dynamic target sensing scheme with (24), (34) and (35)  only needs one beam sweep to obtain the angle, distance  and velocity estimations  for all the dynamic targets, which we term 
\emph{You Only Listen Once} (YOLO for short).

Fig.~7 shows an example of the estimation of  target velocity and distance  with YOLO, in which the target  is one of the four targets  in Fig.~6 and is set as $(120m, 30^\circ,-5m/s)$. According to the frequency of the peak power subcarrier in Fig.~6 and  (24), 
this target angle can be estimated as  $30.0582^\circ$. Then according to  (34) and (35), the velocity and distance estimates of this target can be clearly obtained from Fig.~7 as $-5m/s$ and $120m$, respectively.

\subsection{Multiple YOLO: An Improved Localization Scheme}

Although the proposed YOLO algorithm is theoretically valid, its performance under noise may not be  satisfactory.
The basic idea  of target distance and velocity estimation in YOLO  is to use the phase differences of the $2\ddot M+1$  sidelobe subcarriers with different frequencies and $N_s$ continuous OFDM symbols in only one  beam sweep.
However, it can be seen from (19) that the power of the $k^*$-th target's  sidelobe subcarriers is smaller than that of the peak power subcarrier.
Hence the other $K-1$ targets will cause relatively higher  interference to the
$k^*$-th target's sidelobe subcarriers than its peak power subcarrier,
which may lead to a non-zero error floor for distance and velocity estimation.
Conversely,
the peak power subcarrier contains more pure and accurate target distance and velocity information than  sidelobe subcarriers.

\begin{figure}[!t]
\centering
\includegraphics[width=90mm]{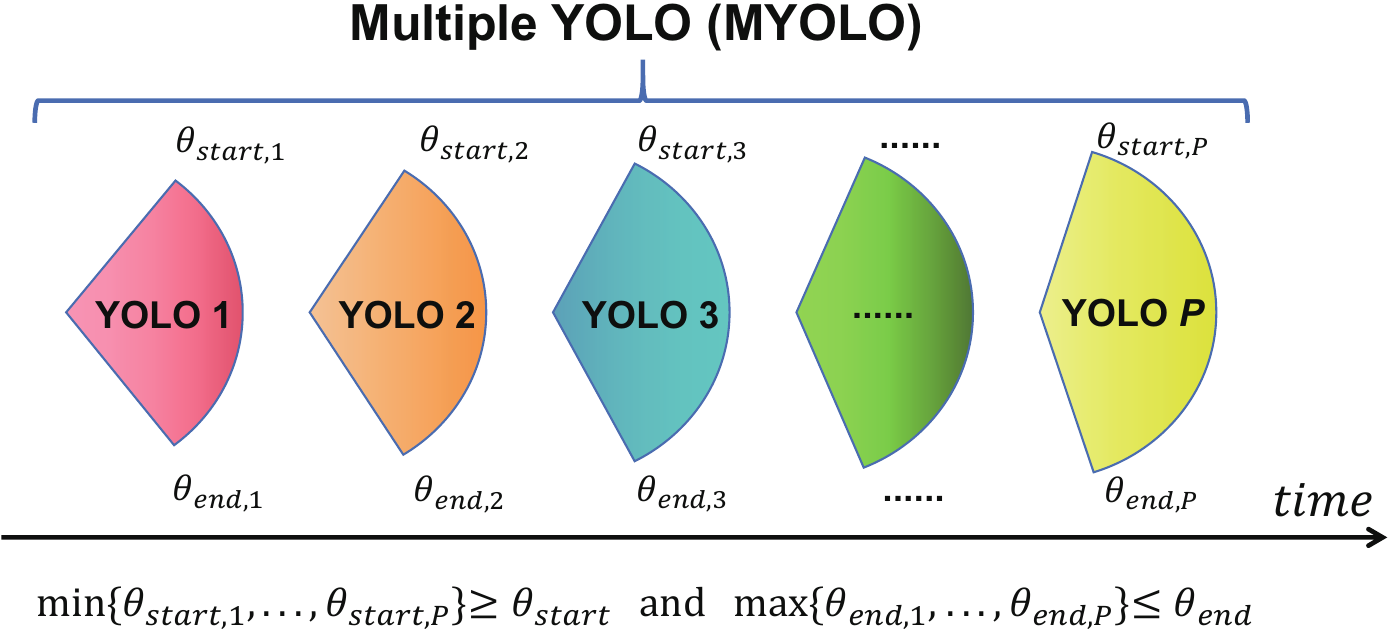}
\caption{Schematic diagram of MYOLO scheme.}
\label{fig_1}
\end{figure}

\begin{figure*}[!t]
\centering
\subfloat[]{\includegraphics[width=75mm]{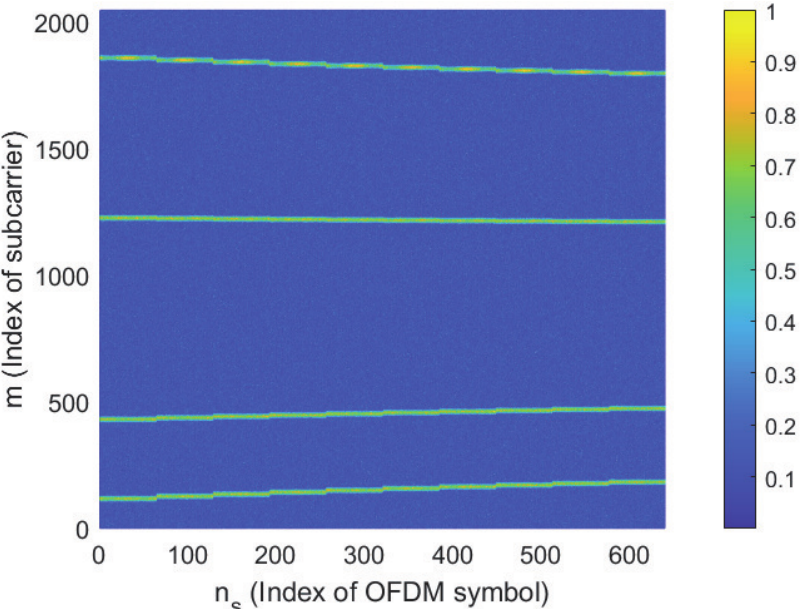}%
\label{fig_first_case}}
\hfil
\subfloat[]{\includegraphics[width=75mm]{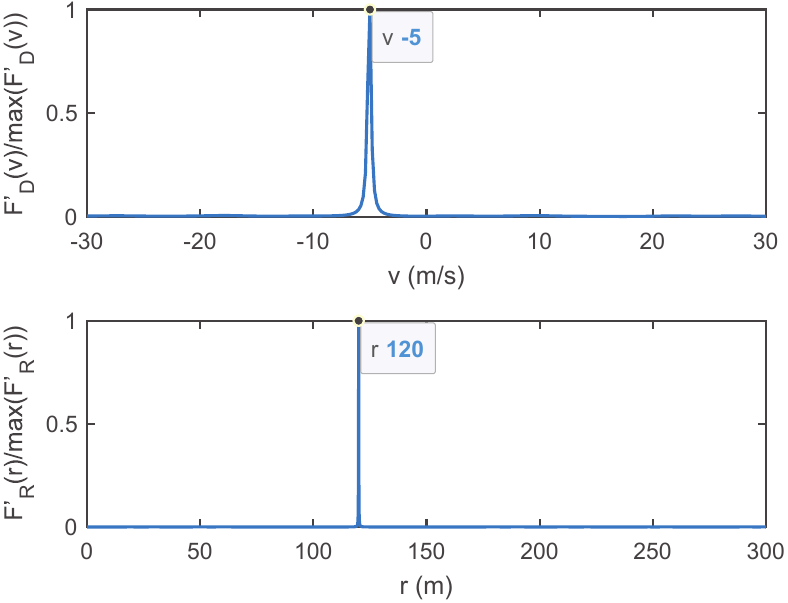}%
\label{fig_second_case}}
\caption{(a) An example of the  power map in MYOLO, in which the four targets are set as $(50m,50^\circ,10m/s)$, $(120m,30^\circ,-5m/s)$, $(200m,-10^\circ,15m/s)$ and $(80m,-45^\circ,8m/s)$.
(b) An example of velocity and distance estimation spectra in MYOLO,
which refers to the dynamic  target with parameter $(120m,30^\circ,-5m/s)$. SNR $=0$ dB,
$N=128$, $M=2048$ and $P=10$.}
\label{fig_sim}
\end{figure*}

Hence we consider listening \emph{more times} to improve the sensing performance.
However, directly repeating YOLO for many times cannot  improve the performance much, because of the interference of the  sidelobe subcarriers in YOLO.
We thus design a new method based on $P$-times beam sweeping with different CBS-Beamforming ranges to realize higher precision target sensing as shown in Fig.~8,
in which we only retain the information of the peak power subcarrier while  discarding  the  sidelobe subcarriers.
By doing this, we set  every time beam sweeping as a YOLO but with different beam squint range.
Specifically,  we set the start angle as $\theta_{start,p}$ and the end angle as
$\theta_{end,p}$ for the $p$-th beam sweep,  where $p=1,2,...,P$.
Note that there should be  $\min\{\theta_{start,1},\theta_{start,2},...,\theta_{start,P}\}\geq\theta_{start}$ and
$\max\{\theta_{end,1},\theta_{end,2},...,\theta_{end,P}\}\leq\theta_{end}$,
such that each  beam sweep can cover the whole sensing range.

Within each beam sweep, the BS can always detect $K$ peak power subcarriers in the echo power spectrum matrix $\mathbf{G}_p$, whose frequencies are denoted by 
 $\{f^{d,1^*,p}_{m},f^{d,2^*,p}_{m},...,f^{d,K^*,p}_{m}\}$.
The angle estimates  of  the $K$ targets can be calculated through (24) as
 $\{\hat{\theta}_{1^*,p},\hat{\theta}_{2^*,p},...,\hat{\theta}_{K^*,p}\}$.
In addition, the echo signal after mean phase cancellation of the peak power subcarrier $f^{d,k^*,p}_{m}$   within the $n_s$-th OFDM symbol time can be written as $\widetilde{y}^{d,k^*,p}_{n_s,m}$, where
${\rm arg}\{\widetilde{y}^{d,k^*,p}_{n_s,m}\}\approx \frac{4\pi f_0v_{k^*}n_sT_s}{c}-\frac{4\pi f^{d,k^*,p}_{m} r_{k^*}}{c}$ can be derived from (23). Then the  received signals of all peak power subcarriers within all OFDM symbols can be represented as 
$\{\widetilde{y}^{d,1^*,p}_{0,m},...,\widetilde{y}^{d,1^*,p}_{(N_s-1),m};...;\widetilde{y}^{d,2^*,p}_{0,m},...,\widetilde{y}^{d,2^*,p}_{(N_s-1),m}\}$.

After $P$-times beam sweep,
we can combine the   power spectrum matrices $\left\{ \mathbf{G}_1, \mathbf{G}_2,...,\mathbf{G}_P \right\}$ into a larger two-dimensional matrix,  called the \emph{power map}.
Since the frequencies  of the $P$ peak power subcarriers corresponding to the same target are adjacent,
we can build the following three information vectors or matrix corresponding to the $k^*$-th target as
\begin{align}
\mathbf{f}^d_{k^*} &= [f^{d,k^*,1}_m,f^{d,k^*,2}_m,...,f^{d,k^*,P}_m]^T                                 \in \mathbb{C}^{P\times 1}  \label{1},\\
\bm{\hat{\theta}}^d_{k^*}&=[\hat{\theta}_{k^*,1},\hat{\theta}_{k^*,2},...,\hat{\theta}_{k^*,P}]^T \in \mathbb{C}^{P\times 1} \label{2},\\
\mathbf{\widetilde{Y}}^d_{k^*} &=
\begin{bmatrix}
\widetilde{y}^{d,k^*,1}_{0,m} & \cdots &\widetilde{y}^{d,k^*,P}_{0,m}   \\
	\vdots   & \ddots   & \vdots  \\
\widetilde{y}^{d,k^*,1}_{(N_s-1),m} & \cdots &\widetilde{y}^{d,k^*,P}_{(N_s-1),m} \\
\end{bmatrix} \in \mathbb{C}^{N_s\times P} \label{3}.
\end{align}
Fig.~9(a) shows an example of the  power map, where the horizontal axis and the vertical axis are the indices of OFDM symbols and subcarriers respectively, and the dynamic targets' parameters are consistent with the settings in Fig.~6.
It is seen that the peak power subcarriers corresponding to the same target
approximately formulate  a line.
Moreover, the number of the lines is $4$, which indicates
 that there are four dynamic targets within the sensing range.

For angle estimation, we can
 take the average value of $\bm{\hat{\theta}}^d_{k^*}$ as the final angle estimation  of the $k^*$-th target, i.e.
\begin{equation}
\begin{split}
\begin{aligned}
\label{deqn_ex1a}
\hat{\theta}_{k^*} = \frac{1}{P}\sum_{p=1}^{P}\hat{\theta}_{k^*,p}.
\end{aligned}
\end{split}
\end{equation}
In addition,
note that the frequencies of the $P$ peak power subcarriers corresponding to the $k^*$-th target  are different thanks to the slightly different range of beam squint  at each time beam sweeping.
Hence we can still estimate the distances and velocities of the dynamic targets using the phase differences of the echo signals from different frequency subcarriers within different OFDM symbols. 
Specifically,
 we  normalize each element in $\mathbf{\widetilde{Y}}^d_{k^*}$ to obtain a new peak power subcarrier echo signal matrix $\mathbf{\bar{Y}}^d_{k^*}$, which satisfies $[\mathbf{\bar{Y}}^d_{k^*}]_{i,j}=\frac{[\mathbf{\widetilde{Y}}^d_{k^*}]_{i,j}}{|[\mathbf{\widetilde{Y}}^d_{k^*}]_{i,j}|}$.
Since the phase of $\mathbf{\bar{Y}}^d_{k^*}$ in  (38) is still linearly uniform in the Doppler dimension,
the Doppler steering vector in the $N_s$ dimension is still defined as
$\mathbf{k}_{D,k^*}(v)=[1,e^{j\frac{4\pi f_0vT_s}{c}},...,e^{j\frac{4\pi f_0vT_s}{c}(N_s-1)}]^T \in \mathbb{C}^{N_s\times 1} $.  However, because the phase of $\mathbf{\bar{Y}}^d_{k^*}$  is non-uniform in the distance dimension, it is necessary to define the distance steering vector corresponding to the $k^*$-th target as $\mathbf{k}_{R,k^*}(r)=[1, e^{-j\frac{4\pi r  }{c}(f_m^{d,k^*,2}-f_m^{d,k^*,1})},...,e^{-j\frac{4\pi r  }{c}(f_m^{d,k^*,P}-f_m^{d,k^*,1})}    ]^T \in \mathbb{C}^{P \times 1}$.
 Then $\mathbf{\bar{Y}}^d_{k^*}$ and its  transpose   can be  represented as $\mathbf{\bar{Y}}^d_{k^*} = e^{-j\frac{4\pi f_m^{d,k^*,1} r_{k^*}}{c}} \mathbf{k}_{D,k^*}(v_{k^*}) 
\mathbf{k}^T_{R,k^*}(r_{k^*}) $ and $(\mathbf{\bar{Y}}^d_{k^*})^T = e^{-j\frac{4\pi f_m^{d,k^*,1} r_{k^*}}{c}} \mathbf{k}_{R,k^*}(r_{k^*}) 
\mathbf{k}^T_{D,k^*}(v_{k^*}) $.
Similar to  (28) to (35), we   calculate the covariance matrix of the peak power subcarrier signals as $\mathbf{R}^{X'}_{D,k^*}$ and $\mathbf{R}^{X'}_{R,k^*}$,  perform eigenvalue decomposition to obtain $\mathbf{U}'_{D,k^*}$, $\mathbf{\Sigma}'_{D,k^*}$, 
$\mathbf{U}'_{R,k^*}$ and $\mathbf{\Sigma}'_{R,k^*}$, and extract the noise subspaces  as
$\mathbf{U}^{N'}_{D,k^*}$ and $\mathbf{U}^{N'}_{R,k^*}$. 
Then  the velocity and distance estimates of the $k^*$-th target can be obtained as
\begin{equation}
\begin{split}
\begin{aligned}
\label{deqn_ex1a}
\!\!\!\!\hat{v}_{k^*}\!&=\!\mathop{\arg\max}\limits_{v}F'_{D,k^*}(v)\\&=
\mathop{\arg\max}\limits_{v} \frac{1}{\mathbf{k}^H_{D,k^*}(v)\mathbf{U}^{N'}_{D,k^*}(\mathbf{U}^{N'}_{D,k^*})^H\mathbf{k}_{D,k^*}(v)},
\end{aligned}
\end{split}
\end{equation}
\begin{equation}
\begin{split}
\begin{aligned}
\label{deqn_ex1a}
\!\!\!\!\hat{r}_{k^*}\!&=\!\mathop{\arg\max}\limits_{r}F'_{R,k^*}(r)\\&=
\mathop{\arg\max}\limits_{r}  \frac{1}{\mathbf{k}^H_{R,k^*}(r)\mathbf{U}^{N'}_{R,k^*}(\mathbf{U}^{N'}_{R,k^*})^H\mathbf{k}_{R,k^*}(r)}.
\end{aligned}
\end{split}
\end{equation}

Different from YOLO, the steps from (39) to (41) utilize the correlation among the peak power subcarriers corresponding to the same target in $P$-times beam sweeping to sense the target, which is termed \emph{Multiple-YOLO} (MYOLO for short).
MYOLO greatly reduces the interference among multiple targets, which not only improves the sense accuracy, but also eliminates the need to allocate $2\ddot M+1$ sidelobe subcarriers for each target as did in YOLO.
Hence  MYOLO can support the sensing of more targets than YOLO under the same configuration. Meanwhile, the selection of $P$ provides a tradeoff between the time required for sensing and the sensing performance.

For example, Fig.~9(b) shows the $v$-$F'_{D}(v)$ and  $r$-$F'_{R}(r)$  curves of searching the velocity and  distance for the dynamic target with $(120m, 30^\circ,-5m/s)$.
 According to the peak value of each search curve, the
estimation results of 
  velocity and   
 distance   for this target can be seen as $-5m/s$ and $120m$.
Comparing Fig.~9(b) and Fig.~7, we can see that the peak of the spectrum search curves in MYOLO are narrower than those of YOLO, which indicates  that MYOLO can better isolate the mutual interference among multiple targets.

\subsection{Complexity analysis}

In the YOLO scheme, the computational complexity mainly comes from the eigenvalue decomposition of $\mathbf{R}^X_{D,k^*}$ and $\mathbf{R}^X_{R,k^*}$. Thus the total computational complexity of the YOLO scheme is $\mathcal{O}\left(N_s^3 + (2\ddot{M}+1)^3\right)$.
Similarly, the computational complexity of the MYOLO scheme mainly comes from the eigenvalue decomposition of $\mathbf{R}^{X'}_{D,k^*}$ and $\mathbf{R}^{X'}_{R,k^*}$. Thus the total computational complexity of the MYOLO scheme can be expressed as 
$\mathcal{O}\left(N_s^3+P^3\right)$.

\section{Simulation Results}

In our simulations, we set
 the number of antennas as $N=128$, the lowest carrier frequency as $f_0$ = 220 GHz and  $d=\frac{1}{2}\lambda$.
We assume that the required sensing range of the BS is $[\theta_{start},\theta_{end}]=[60^\circ,-60^\circ]$.
The noise is assumed to obey the complex Gaussian distribution with mean $\mu=0$ and  variance $\sigma_n^2=1$.
The root  mean square errors (RMSEs) of angle estimation,  distance estimation,  and velocity estimation are defined as
${\rm RMSE}_\theta=\sqrt{\frac{\sum _{i=1}^{Count}(\hat{\theta}_{s(i)}-\theta_{s})^2}{Count}}$, ${\rm RMSE}_r=\sqrt{\frac{\sum _{i=1}^{Count}(\hat{r}_{s(i)}-r_{s})^2}{Count}}$,  and ${\rm RMSE}_v=\sqrt{\frac{\sum _{i=1}^{Count}(\hat{v}_{s(i)}-v_{s})^2}{Count}}$,
where $Count$ is the number of  repeated experiments,
the true parameters of the dynamic target are 
 $(r_{s},\theta_{s},v_s)$, and $(\hat{r}_{s(i)},\hat{\theta}_{s(i)},\hat{v}_{s(i)})$ are the estimated parameters of the target.

\subsection{Dynamic Target Sensing Performance}

Assume that the transmission bandwidth is $W=1$ GHz,  the number of OFDM symbols is $N_s=32$, and the number of subcarriers is $M+1=2048$.
In the YOLO scheme, we set the range of beam squint as
$[\theta_{start},\theta_{end}]=[60^\circ,-60^\circ]$ and the sidelobe window width as $2\ddot M+1=21$. In the MYOLO scheme, 
we set $P=8$, $\theta_{start,p}=60^\circ, 61^\circ, 62^\circ,...,67^\circ$,
and $\theta_{end,p}=-\theta_{start,p}=-60^\circ, -61^\circ, -62^\circ,...,-67^\circ$.
Fig.~10 shows the  RMSEs of dynamic target sensing  versus signal-to-noise ratio (SNR).

\begin{figure*}[!t]
\centering
\subfloat[]{\includegraphics[width=60mm]{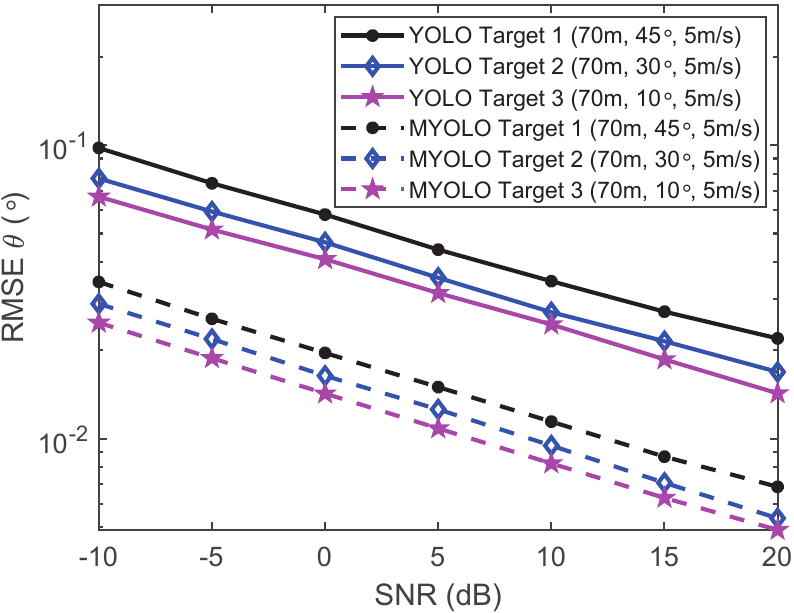}%
\label{fig_first_case}}
\hfil
\subfloat[]{\includegraphics[width=60mm]{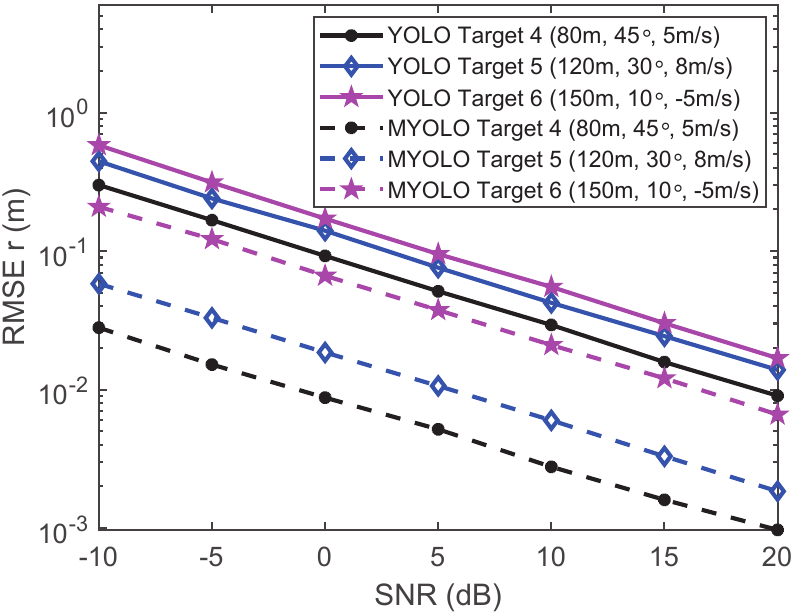}%
\label{fig_first_case}}
\hfil
\subfloat[]{\includegraphics[width=60mm]{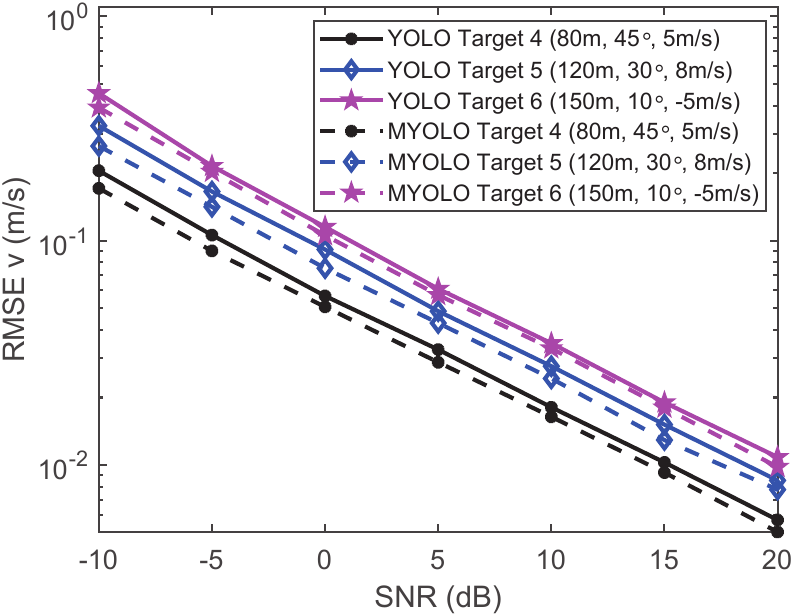}%
\label{fig_second_case}}
\caption{(a) Angle sensing RMSE performance results. (b) Distance sensing  RMSE performance results.
(c) Velocity sensing RMSE performance results.}
\label{fig_sim}
\end{figure*}

\begin{figure*}[!t]
\centering
\subfloat[]{\includegraphics[width=60mm]{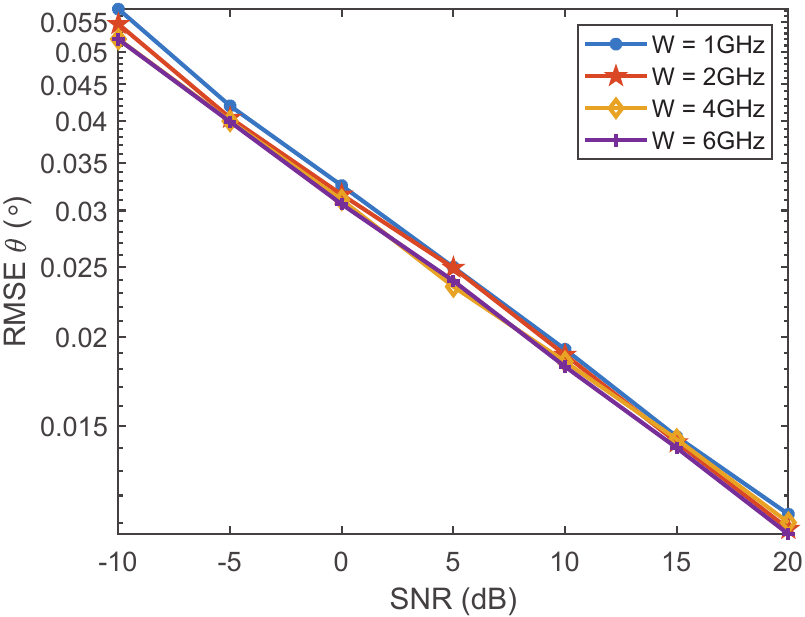}%
\label{fig_first_case}}
\hfil
\subfloat[]{\includegraphics[width=60mm]{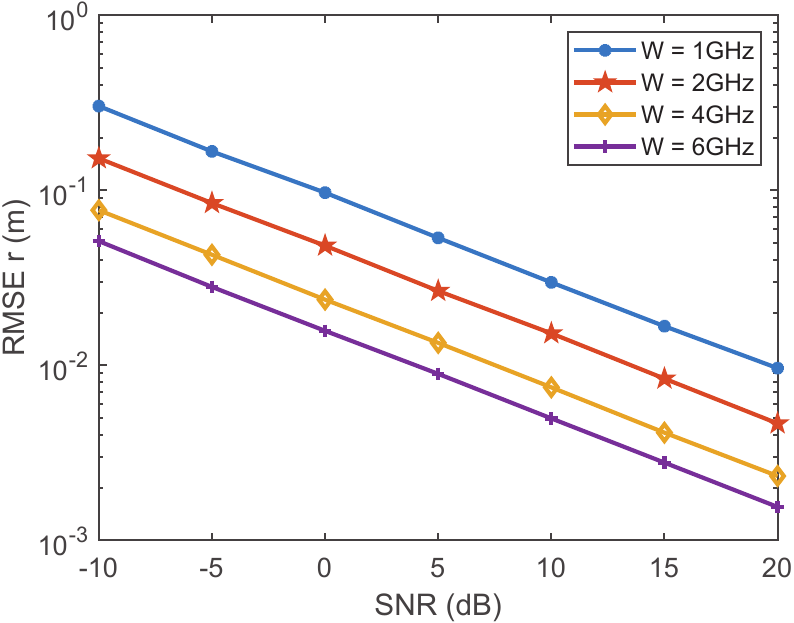}%
\label{fig_first_case}}
\hfil
\subfloat[]{\includegraphics[width=60mm]{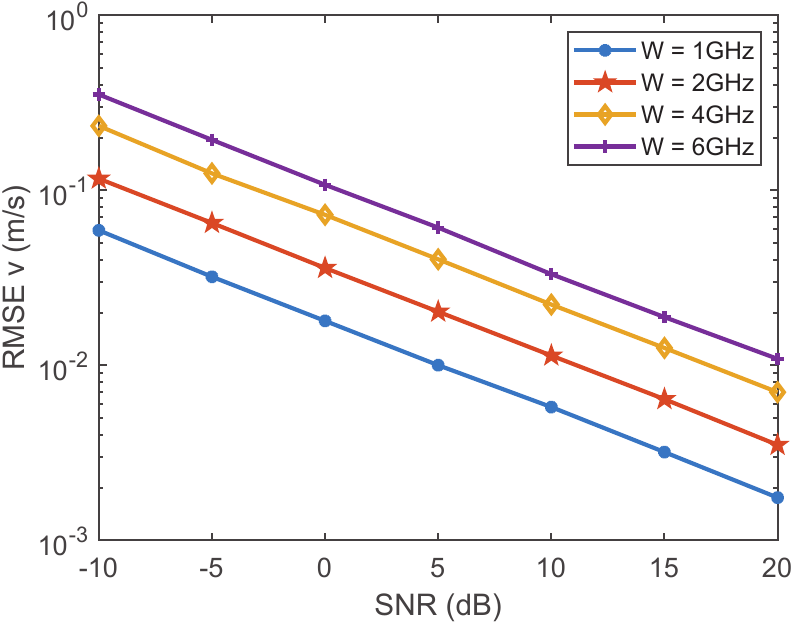}%
\label{fig_second_case}}
\caption{(a) The angle sensing RMSE for different bandwidths.
(b) The distance sensing RMSE for different bandwidths. 
(c) The velocity sensing RMSE for different bandwidths.  The single dynamic target is set as $(200m,10^\circ,5m/s)$. }
\label{fig_sim}
\end{figure*}

It can be seen from Fig.~10(a) that
$\rm RMSE_\theta$ keeps decreasing as SNR increases.
For the YOLO scheme, the average $\rm RMSE_\theta$ is $0.08^\circ$ when  SNR is $-10$ dB, and reaches $0.02^\circ$ when  SNR increases to $15$ dB.
Comparing the curves corresponding to different targets, we find that the closer the target is to $0^\circ$, the lower the $\rm RMSE_\theta$ will be.
This phenomenon is due to the non-uniformity of the angle distribution of the beam squint  in  CBS-beamforming as shown in Fig.~4(b).
It can be seen from  (12) that there are more subcarriers distributed around $0^\circ$, and thus the angle spacing between adjacent subcarriers is smaller,
which means that the angle sensing error will also be smaller around $0^\circ$.
In addition, by comparing the solid line and the dotted line, we see that  MYOLO has better angle sensing performance than YOLO.

Fig.~10(b) and Fig.~10(c) show the ${\rm RMSE}_r$  and ${\rm RMSE}_v$ for different  targets.
It is seen from them that
both the ${\rm RMSE}_r$ and  ${\rm RMSE}_v$ decrease as the SNR increases, and the ${\rm RMSE}_r$  and ${\rm RMSE}_v$ of the target
is larger if the target is father away.
More importantly,  by comparing the solid line and the dotted line, we can see that when other conditions are the same, the MYOLO scheme has higher distance estimation accuracy and velocity estimation accuracy than the YOLO scheme.
This is because MYOLO reduces the mutual interference among multiple targets at the expense of more sensing time.

\subsection{Influence of Bandwidth $W$}

In a practical  Terahertz ISAC system,  only partial spectrum resources are used for sensing, while  other spectrum resources are used for communications.
Therefore, we consider changing $W$ to explore the impact of bandwidth.
Fig.~11 shows the sensing RMSEs of the YOLO scheme when a single dynamic target is set as $(200m,10^\circ,5m/s)$, while the bandwidth varies.
As $W$ decreases, the ${\rm RMSE}_\theta$  is almost unchanged, while the ${\rm RMSE}_r$  gradually increases.
This is consistent with the basic intuition that
the accuracy of distance estimation in radar system increases with the increase of system bandwidth, while the angle estimation accuracy is almost independent of the bandwidth.
In addition, due to the fixed number of subcarriers $M+1$ in the experiment, as the bandwidth $W$ increases, the frequency interval between subcarriers $\Delta f$ gradually increases, leading to a decrease in velocity resolution, which explains the phenomenon of 
${\rm RMSE}_v$ increasing as $W$ increasing in Fig.~11(c).

\begin{figure*}[!t]
\centering
\subfloat[]{\includegraphics[width=60mm]{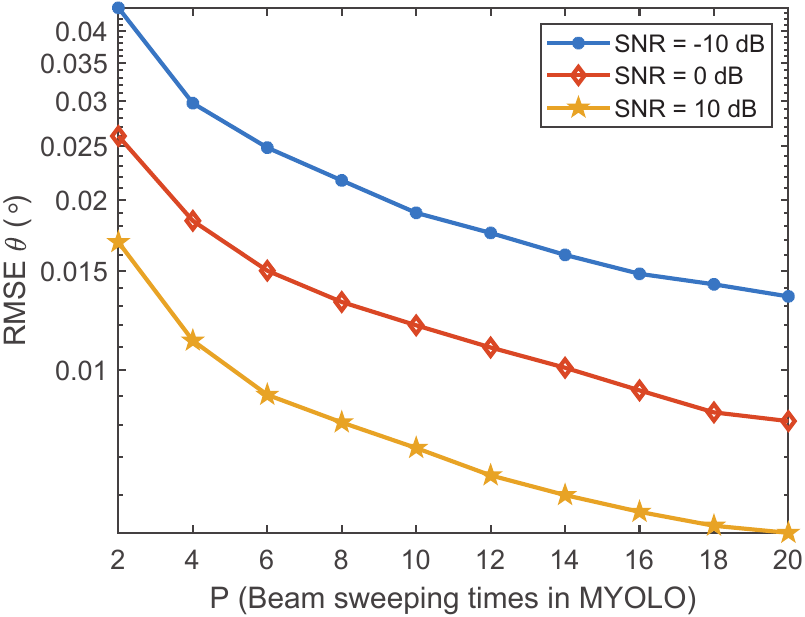}%
\label{fig_first_case}}
\hfil
\subfloat[]{\includegraphics[width=60mm]{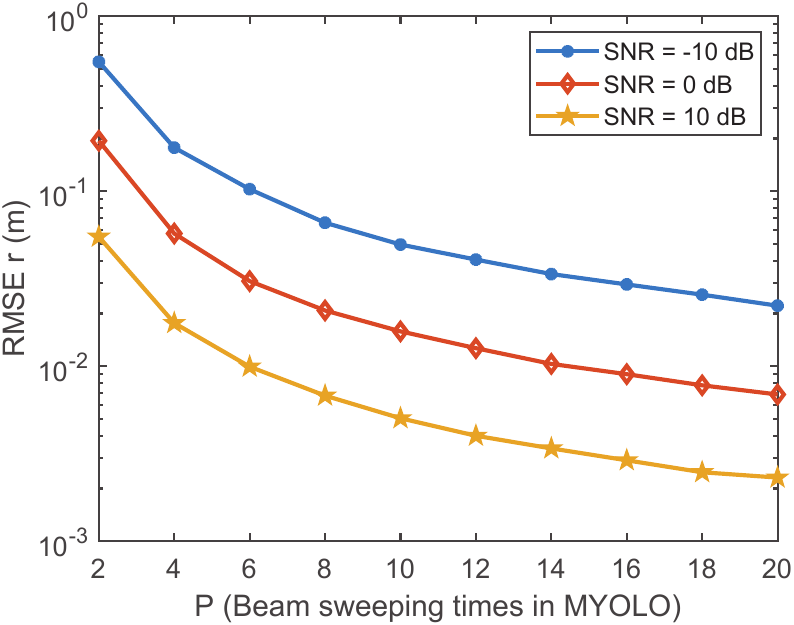}%
\label{fig_first_case}}
\hfil
\subfloat[]{\includegraphics[width=60mm]{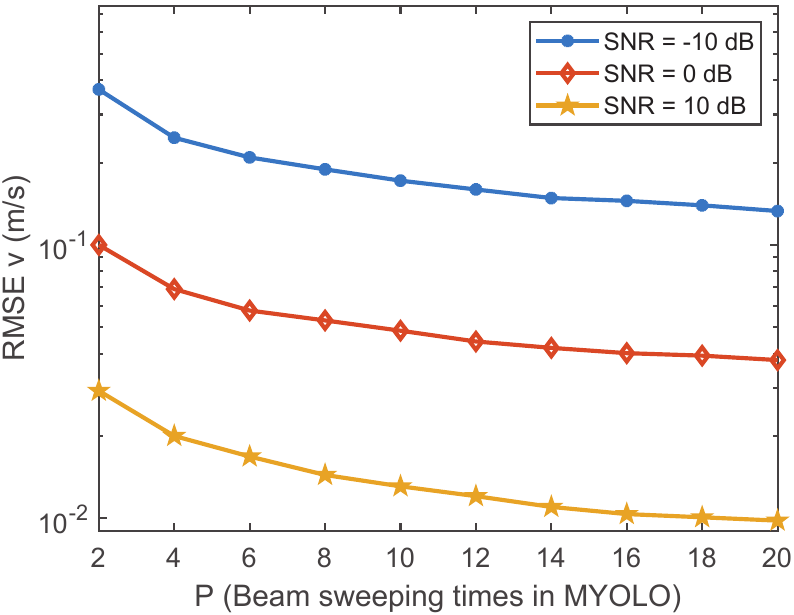}%
\label{fig_second_case}}
\caption{(a) The angle sensing MSE vs. the number of beam sweeps.
(b)The distance sensing MSE vs. the number of beam sweeps. (c)The velocity sensing MSE vs. the number of beam sweeps.}
\label{fig_sim}
\end{figure*}

\subsection{Influence of Beam Sweeping Times $P$}

We study the effect of changing $P$ in  the MYOLO sensing scheme.
It can be seen from Fig.~12(a) that the ${\rm RMSE}_\theta$ decreases
with the increase of $P$. When $P=4$ and SNR $=10$ dB, the ${\rm RMSE}_\theta$ is $0.011^\circ$, while when $P$ increases to $12$, the ${\rm RMSE}_\theta$ decreases to $0.007^\circ$.
This is mainly because MYOLO's angle estimation method, formula (39), takes the mean value of the angle estimation results obtained from $P$ times YOLO scheme. Therefore, the angle estimation accuracy will improve as $P$ increases.

It can be seen from Fig.~12(b) that the ${\rm RMSE}_r$ decreases
with the increase of $P$. 
When $P=4$ and SNR $=10$ dB, the ${\rm RMSE}_r$ is $0.017m$, while
when $P$ increases to $12$, the ${\rm RMSE}_r$ decreases to $0.005m$.
This is mainly because in the MYOLO scheme, the distance array represented by the distance steering vector increases with the increase of $P$, and the larger distance arrays are more conducive to distance estimation.

\begin{figure*}[!t]
\centering
\subfloat[]{\includegraphics[width=60mm]{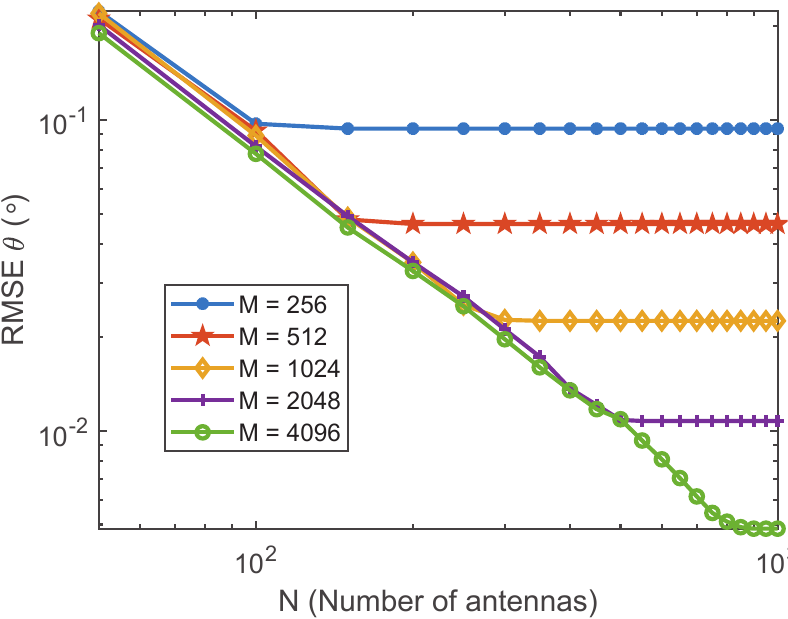}%
\label{fig_first_case}}
\hfil
\subfloat[]{\includegraphics[width=60mm]{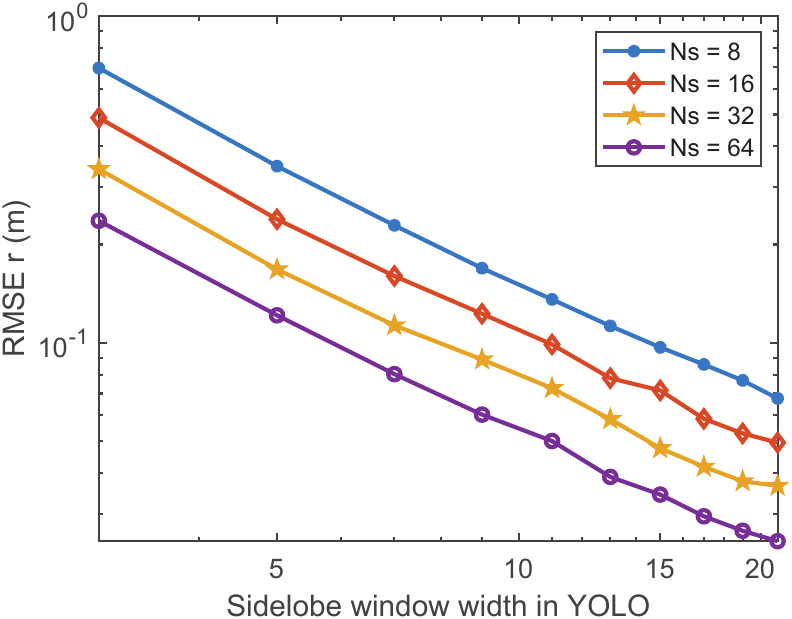}%
\label{fig_first_case}}
\hfil
\subfloat[]{\includegraphics[width=60mm]{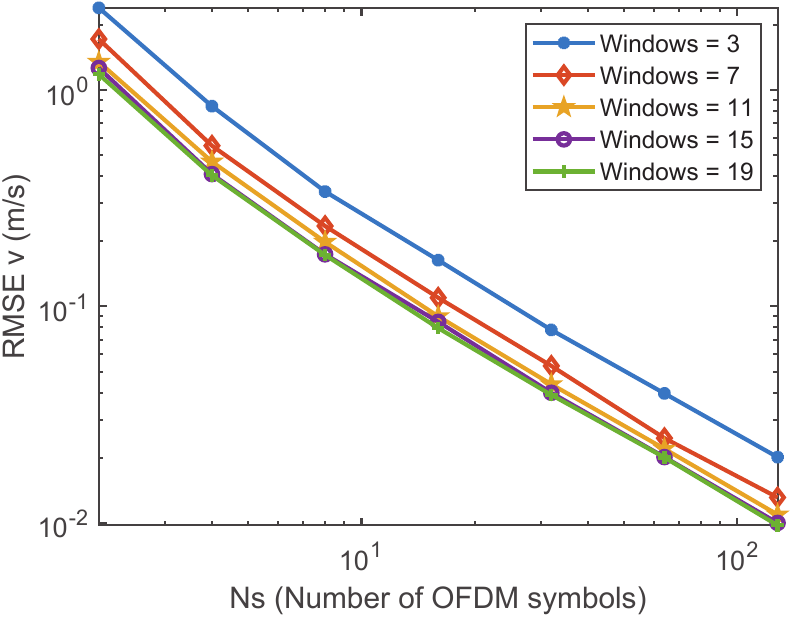}%
\label{fig_second_case}}
\caption{(a) Relationship between the angle sensing RMSE and the number of antennas $N$ and the   number of subcarriers $M$.
(b)  Relationship between the distance sensing RMSE and the sidelobe window width $2\ddot M+1$ and the number of  symbols $N_s$. (c) Relationship between  the velocity sensing RMSE and the  number of  symbols $N_s$ and the sidelobe window width $2\ddot M+1$.}
\label{fig_sim}
\end{figure*}

It can be seen from Fig.~12(c) that the ${\rm RMSE}_v$ decreases
with the increase of $P$. 
When $P=4$ and SNR $=10$ dB, ${\rm RMSE}_v$ is $0.020m/s$, while 
when $P$ increases to $12$, the ${\rm RMSE}_v$ decreases to $0.012m/s$. This is mainly because in the MYOLO scheme, the number of observations on the velocity array represented by the velocity steering vector increases with the increase of $P$, and more observations make the covariance matrix calculation of the velocity array more accurate, thereby improving the accuracy of velocity estimation.

Since increasing $P$  requires more time to complete sensing, 
there is a tradeoff between  the sensing accuracy and the acceptable sensing time.

\subsection{Influence of Other System Parameters}

Fig.~13 shows the impact of some other system parameters on sensing performance. 
Fig.~13(a) shows the variation  ${\rm RMSE}_\theta$ with the number of antennas and subcarriers.
It is seen that the ${\rm RMSE}_\theta$  decreases with the increase of $N$. According to the theory of array signal processing, the beam width of the antenna array is inversely proportional to $2N$.
Therefore, by increasing  $N$, the BS can concentrate the energy in a narrower width to improve the angle sensing performance.
However, it is seen that when  $N$ is large enough, ${\rm RMSE}_\theta$ will converge to a non-zero error floor, which will decrease with the increase of the number of subcarriers  $M$.
The reason for this is that when $M$ is small, the distribution interval between subcarriers is large, which limits the angle sensing performance of the beams with very narrow beam width.

Fig.~13(b) shows the  distance sensing RMSE in the YOLO scheme versus the sidelobe  window width $2\ddot M+1$ and the number of OFDM symbols $N_s$.  It can be seen  from  the figure that the distance RMSE  gradually decreases as the width of the sidelobe window increases. This is because a larger sidelobe window width means a larger distance steering array, making distance sensing more accurate. In addition, it can  be found that increasing the number of OFDM symbols can also improve the accuracy of distance sensing.

Fig.~13(c) shows the  velocity sensing RMSE in the YOLO scheme versus the number  of OFDM symbols $N_s$ and the sidelobe window width $2\ddot M+1$. From this figure, it can be seen that using more OFDM symbols can improve the accuracy of velocity sensing.
In addition, increasing the width of the sidelobe window can also improve the accuracy of velocity sensing to a certain extent. But as the window width increases, the improvement  gradually decreases.

\section{Conclusions}

In this paper, we have proposed the use of the 
 beam squint effect in massive MIMO Terahertz wideband communications systems to realize non-cooperative dynamic target sensing.
Specifically,
we have constructed a wideband channel model for  echo signals,
and proposed a  beamforming scheme that controls the range of beam squint by adjusting the values of PSs and TTDs, such that different subcarriers can point to different directions in a planned way.
The echo signals of different subcarriers will carry target information in different directions, based on which  the targets' angles can be estimated through a sophisticatedly designed algorithm.
Moreover,
we have proposed a supporting method based on extended array signal estimation, which utilizes the phase changes of different frequency subcarriers within different OFDM symbols to estimate the distance and velocity of dynamic targets.
Interestingly,
the proposed method only needs the BS to transmit and receive the signals once to realize full spatial target localization, which means that \emph{You Only Listen Once} (YOLO). 
In order to further improve the accuracy of sensing, we have proposed a high-precision sensing scheme based on multiple applications of  YOLO, which utilized the mean estimation, non-uniform array signal estimation, and uniform array signal estimation to realize high-precision estimation of dynamic target angle, distance, and velocity, respectively.
Compared with the traditional ISAC method that requires multiple beam sweeps, the proposed one can greatly reduce the sensing time overhead.
Simulation results have demonstrated the effectiveness of the proposed scheme.

\section*{APPENDIX A \\ Proof of (15)}
The echo signal after clutter filtering    is $\widetilde{y}_{n_s,m}$, which is given by  (14) and the phase of the additive term in $\widetilde{y}_{n_s,m}$ is $2\pi (\phi_{n}+\widetilde{f}_{m}t_{n}+\frac{f_m}{c}nd\sin\theta_k)$. We set $\phi_{n}=-\frac{f_0nd\sin\theta_{start}}{c}$ and $t_{n}=-\frac{\phi_{n}}{W}-\frac{(f_0+W)nd\sin\theta_{end}}{Wc}$.
Then we have $2\pi (\phi_{n}+\widetilde{f}_{m}t_{n}+\frac{f_m}{c}nd\sin\theta_k)= -\frac{2\pi dn}{Wc}[(W\!-\!\widetilde{f}_{m})f_0\sin\theta_{start}+(W\!+\!f_0)\widetilde{f}_{m}\sin\theta_{end}-Wf_m\sin\theta_k]=-n\beta(f_m,\theta_k)$,
where $\beta(f_m,\theta_k)=\frac{2\pi d}{Wc}[(W-\widetilde{f}_{m})f_0\sin\theta_{start}+(W+f_0)\widetilde{f}_{m}\sin\theta_{end}-Wf_m\sin\theta_k]$.
Therefore (14) is simplified as
\begin{equation}
\begin{split}
\begin{aligned}
\label{deqn_ex1a}
 &\widetilde{y}_{n_s,m}\\&\approx \sum_{k=1}^{K}\!\!\left\{\!\!
\frac{\alpha_k}{N}e^{j\phi{(k,n_s,f_m)}}\!\!
\left[\!\!\sum_{n=-\frac{N-1}{2}}^{\frac{N-1}{2}}\!\!\!\!\!\!e^{j2\pi\phi_{n}}
\!e^{j2\pi \widetilde{f}_{m}t_{n}}\!e^{j\frac{2\pi f_m}{c}nd\sin\theta_{k}}\right]^2
\!\!\right\}
\\
&=\sum_{k=1}^{K}\left\{
\frac{\alpha_k}{N} e^{j\phi{(k,n_s,f_m)}}\times
\left[\sum_{n=-\frac{N-1}{2}}^{\frac{N-1}{2}}
e^{-jn\beta(f_m,\theta_k)}
\right]^2\right\}\\
&=\sum_{k=1}^{K}\left\{
\frac{\alpha_k}{N}e^{j\phi{(k,n_s,f_m)}}\left[\frac{\sin(\frac{\beta(f_m,\theta_k)}{2}N)}{\sin(\frac{\beta(f_m,\theta_k)}{2})}\right]^2\right\}.
\end{aligned}
\end{split}
\end{equation}
This completes the proof of (15).

\bibliographystyle{ieeetr}
\bibliography{YOLO.bib}

\begin{IEEEbiography}[{\includegraphics[width=1in,height=1.25in,clip,keepaspectratio]{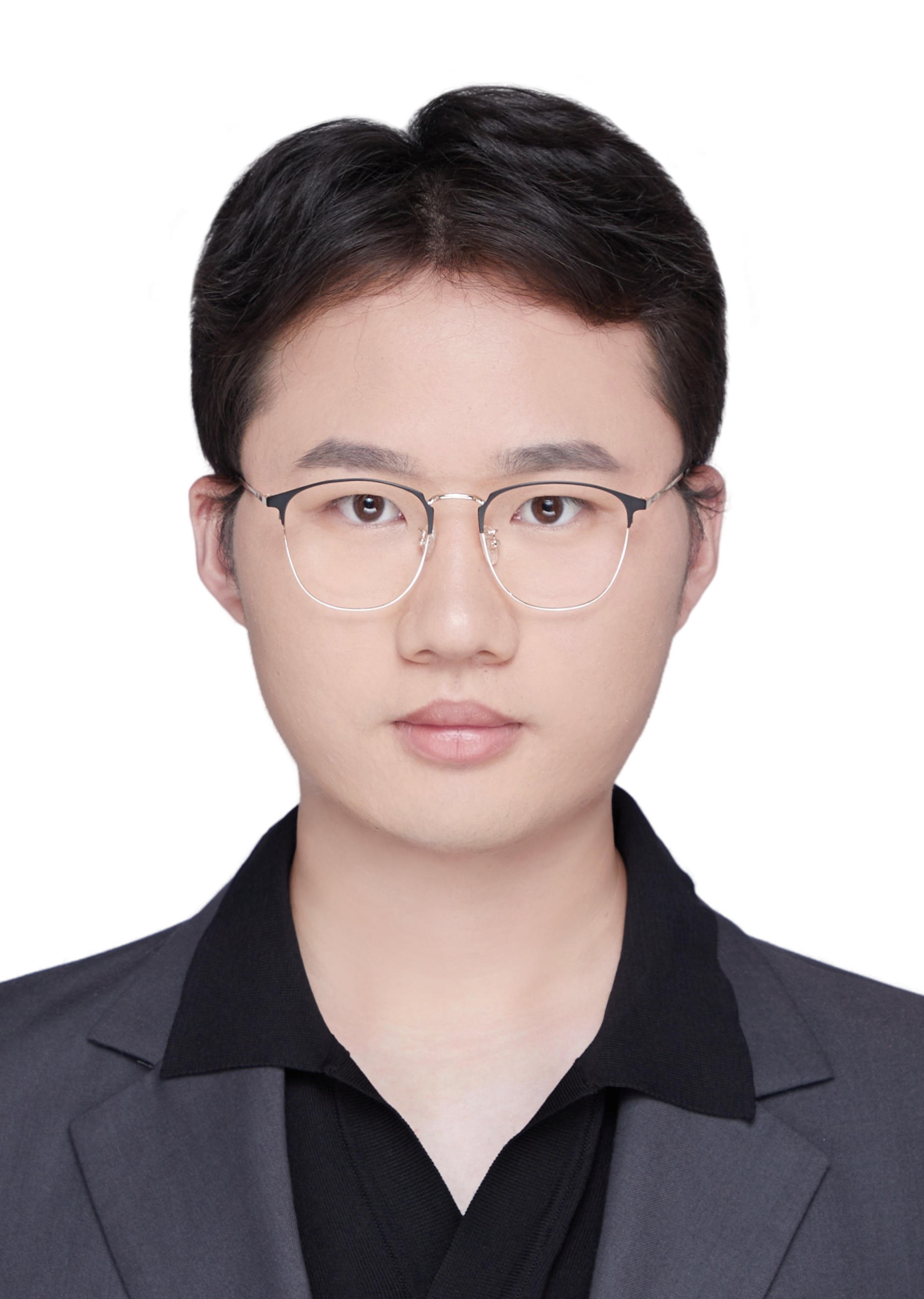}}]{\\Hongliang Luo}
received the B.Eng. degree from Xidian University, Xi'an, China, in 2023.
He is currently working toward the Ph.D. degree with the Department of Automation, Tsinghua University, Beijing, China.

His research interests include integrated sensing and communications,  wireless communications, radar sensing, array signal processing, massive MIMO and beamforimg design.
\end{IEEEbiography}

\begin{IEEEbiography}[{\includegraphics[width=1in,height=1.25in,clip,keepaspectratio]{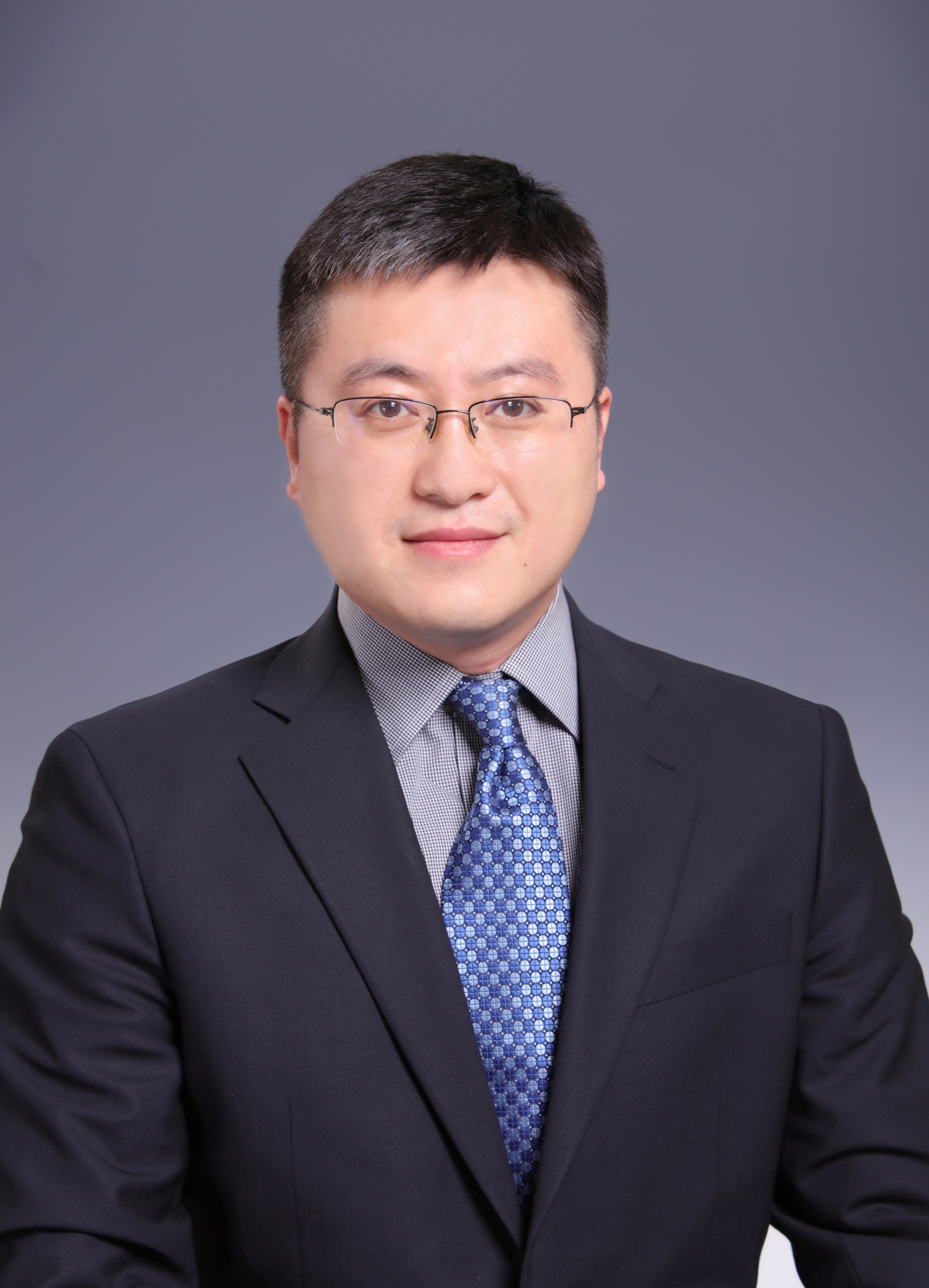}}]{Feifei Gao} 
(Fellow, IEEE)
received the B.Eng. degree from Xi'an Jiaotong University, Xi'an, China in 2002, the M.Sc. degree from McMaster University, Hamilton, ON, Canada in 2004, and the Ph.D. degree from National University of Singapore, Singapore in 2007. Since 2011, he joined the Department of Automation, Tsinghua University, Beijing, China, where he is currently an Associate Professor.

Prof. Gao's research interests include signal processing for communications, array signal processing, convex optimizations, and artificial intelligence assisted communications. He has authored/coauthored more than 200 refereed IEEE journal papers and more than 150 IEEE conference proceeding papers that are cited more than 17000 times in Google Scholar. Prof. Gao has served as an Editor of IEEE Transactions on Wireless Communications, IEEE Journal of Selected Topics in Signal Processing (Lead Guest Editor), IEEE Transactions on Cognitive Communications and Networking, IEEE Signal Processing Letters (Senior Editor), IEEE Communications Letters (Senior Editor), IEEE Wireless Communications Letters, and China Communications. He has also served as the symposium co-chair for 2019 IEEE Conference on Communications (ICC), 2018 IEEE Vehicular Technology Conference Spring (VTC), 2015 IEEE Conference on Communications (ICC), 2014 IEEE Global Communications Conference (GLOBECOM), 2014 IEEE Vehicular Technology Conference Fall (VTC), as well as Technical Committee Members for more than 50 IEEE conferences.
\end{IEEEbiography}

\begin{IEEEbiography}[{\includegraphics[width=1in,height=1.25in,clip,keepaspectratio]{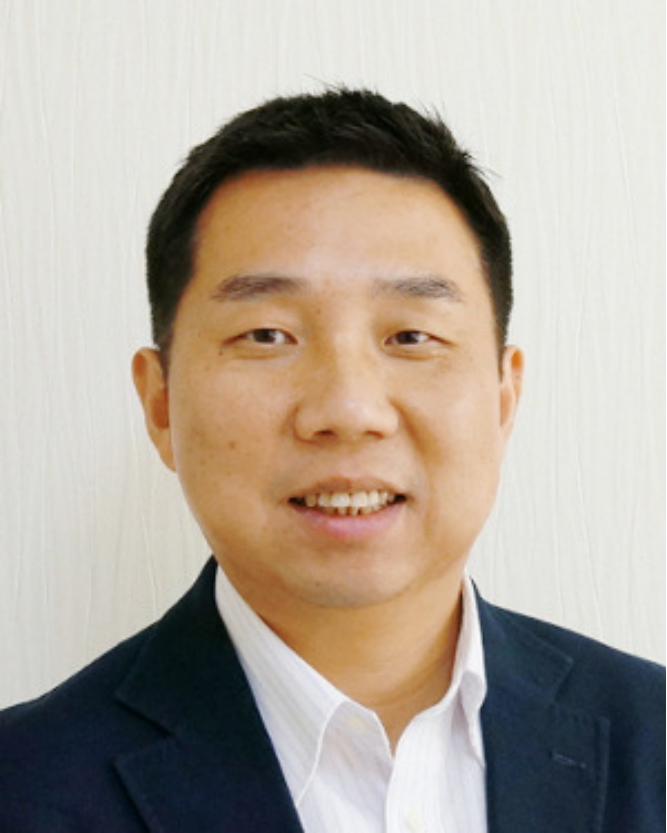}}]{Hai Lin}
	(Senior Member, IEEE) received the B.E.degree from Shanghai Jiaotong University, China,
in 1993, the M.E. degree from University of the Ryukyus, Japan, in 2000, and the Dr.Eng. degree
from Osaka Prefecture University, Japan, in 2005. In 2000, he joined the Graduate School of
Engineering, Osaka Prefecture University (renamed Osaka Metropolitan University, in 2022), where
he is currently a Professor. 

His research interests include signal processing for communications, wireless communications, and statistical signal processing. He served several times as a Technical Program Co-Chair for the Signal Processing for Communications Symposium and the Wireless Communications Symposium of the IEEE International Conference on Communications (ICC) and IEEE Global Communications Conference (GLOBECOM). He was the Chair of the Signal Processing and Communications Electronics Technical Committee, IEEE Communications Society, from 2015 to 2016, and the Chair of the IEEE Communications Society Kansai Chapter from 2022 to 2023. He was an Associate Editor for the IEEE TRANSACTIONS ON WIRELESS COMMUNICATIONS. He is currently serving as an Associate Editor for the IEEE TRANSACTIONS ON COMMUNICATIONS and the IEEE TRANSACTIONS ON VEHICULAR TECHNOLOGY.
\end{IEEEbiography}

\begin{IEEEbiography}[{\includegraphics[width=1in,height=1.25in,clip,keepaspectratio]{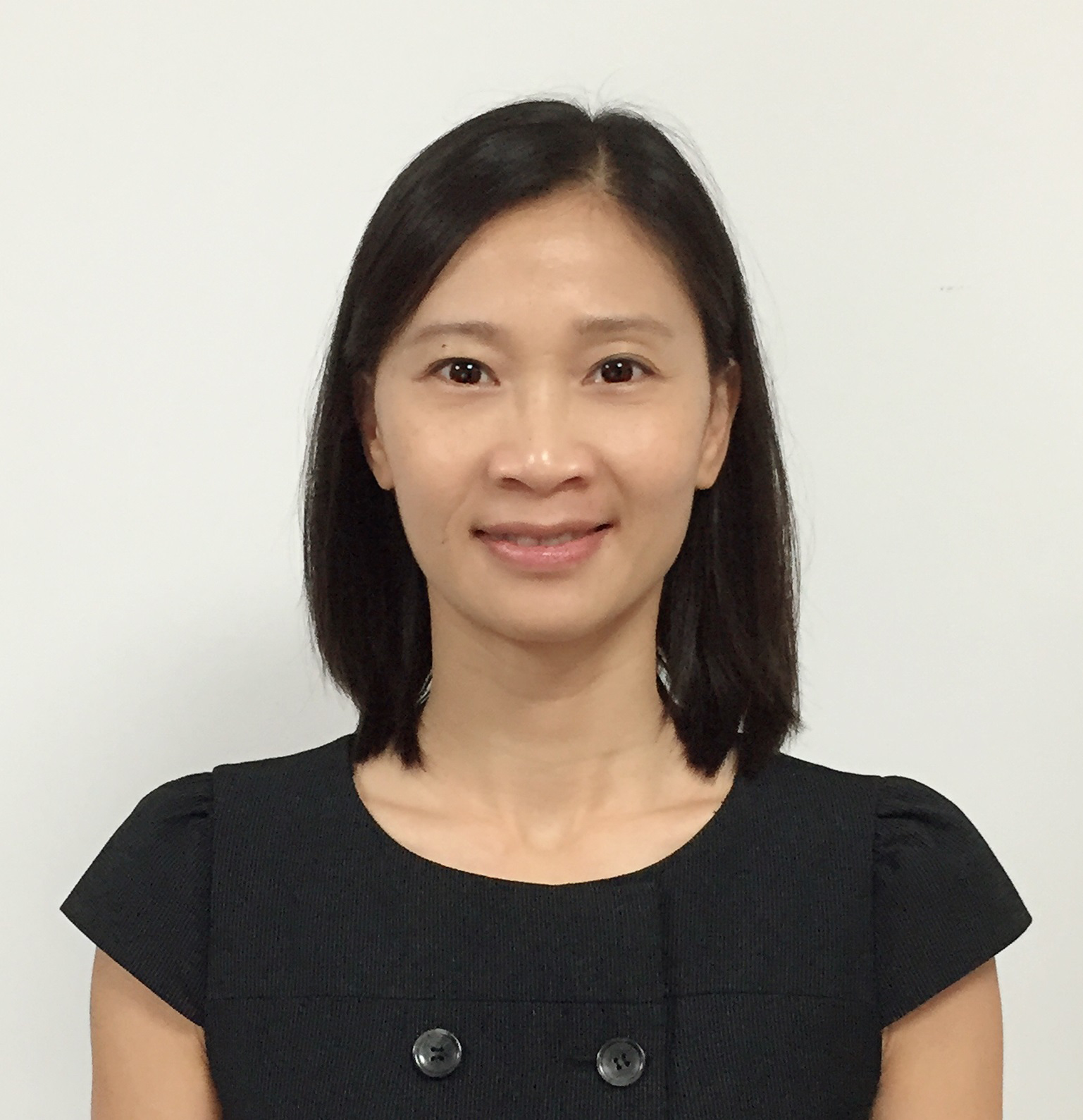}}]{\\Shaodan Ma}
(Senior Member, IEEE) received the double Bachelor’s degrees in science and economics and the M.Eng. degree in electronic engineering from Nankai University, Tianjin, China, in 1999 and 2002, respectively, and the Ph.D. degree in electrical and electronic engineering from The University of Hong Kong, Hong Kong, in 2006. From 2006 to 2011, she was a post-doctoral fellow at The University of Hong Kong. Since August 2011, she has been with the University of Macau, where she is currently a Professor. 
Her research interests include array signal processing, transceiver design, localization, integrated sensing and communication, mmwave communications, massive MIMO, and machine learning for communications. She was a symposium co-chair for various conferences including IEEE ICC 2021, 2019 and 2016, IEEE GLOBECOM 2016, IEEE/CIC ICCC 2019, etc. She has served as an Editor for IEEE Transactions on Wireless Communications (2018-2023), IEEE Transactions on Communications (2018-2023), IEEE Wireless Communications Letters (2017-2022), IEEE Communications Letters (2023), and Journal of Communications and Information Networks (2021-present).
\end{IEEEbiography}

\begin{IEEEbiography}[{\includegraphics[width=1in,height=1.25in,clip,keepaspectratio]{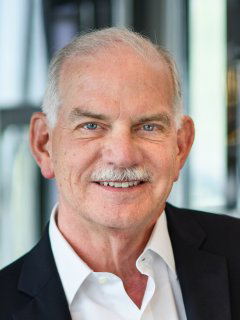}}]{H. Vincent Poor}
(S’72, M’77, SM’82, F’87) received the Ph.D. degree in EECS from Princeton University in 1977.  From 1977 until 1990, he was on the faculty of the University of Illinois at Urbana-Champaign. Since 1990 he has been on the faculty at Princeton, where he is currently the Michael Henry Strater University Professor. During 2006 to 2016, he served as the dean of Princeton’s School of Engineering and Applied Science. He has also held visiting appointments at several other universities, including most recently at Berkeley and Cambridge. 

His research interests are in the areas of information theory, machine learning and network science, and their applications in wireless networks, energy systems and related fields. Among his publications in these areas is the recent book Machine Learning and Wireless Communications.  (Cambridge University Press, 2022). Dr. Poor is a member of the National Academy of Engineering and the National Academy of Sciences and is a foreign member of the Chinese Academy of Sciences, the Royal Society, and other national and international academies. He received the IEEE Alexander Graham Bell Medal in 2017.
\end{IEEEbiography}

\vfill

\end{document}